\documentclass[useAMS,usenatbib]{mn2e}
\usepackage{graphicx}
\usepackage{array,longtable}
\usepackage{natbib}
\usepackage{float}
\usepackage[T1]{fontenc}
\usepackage{ae}
\usepackage{aecompl}
\usepackage{amsmath, amssymb, amsfonts}
\usepackage{multicol}
\def\actaa{Acta. Astronom.} %
\def\aj{AJ}%
\def\apj{ApJ}%
\def\apjl{ApJ}%
\def\aap{A\&A}%
\def\mnras{MNRAS}%

 \title[Milky Way's X-Shaped Bulge]{The X-Shaped Milky Way Bulge in OGLE-III\thanks{Based on observations obtained with the 1.3 m Warsaw telescope at the Las Campanas Observatory of the Carnegie Institution for Science.} Photometry and in N-Body Models}
\author[Nataf et al.]{David M. Nataf$^1$\thanks{Email: david.nataf@anu.edu.au}, Andrzej Udalski$^{2}$,  Jan Skowron$^{2}$, Micha{\l} K. Szyma{\'n}ski$^{2}$, 
\newauthor {Marcin Kubiak$^{2}$, Grzegorz Pietrzy{\'n}ski$^{2,3}$, Igor Soszy{\'n}ski$^{2}$, Krzysztof Ulaczyk$^{2}$, }
\newauthor {{\L}ukasz Wyrzykowski$^{2,4}$, Rados{\l}aw Poleski$^{5,2}$,  E. Athanassoula$^{6}$,  }
\newauthor {Melissa Ness$^{7}$, Juntai Shen$^{8}$,  Zhao-Yu Li$^{8}$}
\vspace*{6pt}\\
$^{1}$Research School of Astronomy and Astrophysics, Australian National University, Canberra, ACT 2611, Australia \\ 
$^{2}$Warsaw University Observatory, Al. Ujazdowskie 4, 00-478 Warszawa,Poland \\ 
$^{3}$Universidad de Concepci{\'o}n, Departamento de Astronomia,
Casilla 160--C, Concepci{\'o}n, Chile \\ 
$^{4}$Institute of Astronomy, University of Cambridge, Madingley Road, Cambridge CB3 0HA, UK \\ 
$^{5}$Department of Astronomy, Ohio State University, 140 W. 18th Ave., Columbus, OH 43210 \\ 
$^{6}$Aix Marseille Universit\'e, CNRS, LAM (Laboratoire d'Astrophysique de Marseille) UMR 7326, 13388, Marseille, France \\ 
$^{7}$Max-Planck-Institut f\"ur Astronomie, K\"onigstuhl 17, 69117 Heidelberg, Germany \\ 
$^{8}$Key Laboratory for Research in Galaxies and Cosmology, Shanghai Astronomical Observatory, Chinese Academy of Sciences, \\ 
80 Nandan Road, Shanghai 200030, China \\ }
\begin{document}
\include{journaldefs}
\date{Accepted ...... Received ...... ; in original form......   }

\pagerange{\pageref{firstpage}--\pageref{lastpage}} \pubyear{2014}
\maketitle
\label{firstpage}
\begin{abstract}
We model the split red clump of the Galactic bulge in OGLE-III photometry, and compare the results to predictions from two N-body models. Our analysis yields precise maps of the brightness of the two red clumps, the fraction of stars in the more distant peak, and their combined surface density. We compare  the observations to predictions from two N-body models previously used in the literature. Both models correctly predict several features as long as one assumes an angle $\alpha_{\rm{Bar}}  \approx 30^{\circ}$ between the Galactic bar's major axis and the line of sight to the Galactic centre. In particular that the fraction of stars in the faint red clump should decrease with increasing longitude. The biggest discrepancies between models and data are in the rate of decline of the combined surface density of red clump stars toward negative longitudes and of the brightness difference between the two red clumps toward positive longitudes, with neither discrepancy exceeding $\sim$25\% in amplitude. Our analysis of the red giant luminosity function also yields an estimate of the red giant branch bump parameters toward these high-latitude fields, and evidence for a high rate ($\sim$25\%) of disk contamination in the bulge at the colour and magnitude of the red clump, with the disk contamination rate increasing toward sightlines further distant from the plane. 
\end{abstract}
\maketitle


\section{Introduction}
\label{sec:Introduction}
The red clump (RC) has been used to trace the spatial structure of the inner Galaxy since the work of \citet{1994ApJ...429L..73S}, who found that the dereddened RC apparent magnitudes were brighter toward positive longitudes. \citet{1994ApJ...429L..73S}, and subsequently and in superior detail, \citet{1997ApJ...477..163S}, thus argued for a bar-shaped structure for the Galactic bulge, with the nearer end of the bar oriented toward positive longitudes with a viewing angle $\alpha_{\rm{Bar}} \approx 25^{\circ}$\footnote{This paper discusses two parameters often denoted as ``$\alpha$", hence the subscripts leading to the parameter names $\alpha_{\rm{Bar}} $ and $\alpha_{\rm{Skew}}$. The former refers to the viewing angle to the Galactic bar, the latter is a parameter in the skew-Gaussian distributions used to model the number counts of red clump stars. } between the bar's major axis and the line of sight between the Sun and the Galactic centre. The advent of larger data surveys inevitably necessitated a more refined geometrical model. \citet{2010ApJ...721L..28N}, in their analysis of OGLE-III photometry \citep{2011AcA....61...83S}, found that the RC brightness distribution shifted from being single-peaked to double-peaked toward sight lines near the Galactic bulge minor axis, far from the plane, with the two peaks having equal or nearly equal $(V-I)$ colour. Concurrently and independently, \citet{2010ApJ...724.1491M}  also found a double RC in OGLE-II \citep{2002AcA....52..217U} and 2MASS data \citep{2006AJ....131.1163S}, which they argued was due to an X-shaped bulge. 

The hypothesis that the bifurcation in the apparent magnitude distribution of the RC was due to an X-shaped bulge was confirmed by \citet{2012ApJ...756...22N}, who compared ARGOS spectroscopic \citep{2013MNRAS.428.3660F}  and photometric data to N-body models \citep{2003MNRAS.341.1179A} and demonstrated a stunning degree of consistency between observations and theoretical predictions. \citet{2012ApJ...756...22N} found that the double RC had to extend from no less than $b=-5^{\circ}$ to $b=-10^{\circ}$, that the two RCs had a radial velocity offset of the same sign and comparable size as dynamical predictions from N-body models,  and that this morphology characterised the kinematics of bulge stars with [Fe/H]  $\gtrsim -0.5$,  in other words the vast majority ($\gtrsim90\%$)  of bulge stars. Simultaneously and independently, \citet{2012ApJ...757L...7L} showed that the split red clump was also predicted by an N-body model previously used to successfully match to Milky Way bulge radial velocity data \citep{2010ApJ...720L..72S}.  Similar and additional kinematic comparisons between N-body model predictions and observations have since been reported by \citet{2013A&A...555A..91V} and \citet{2014MNRAS.438.3275G}. 

This dynamical phenomenon begs detailed characterisation. The orbits that form the X-shaped structure are due to dynamical instabilities \citep{2005MNRAS.358.1477A, 2012ApJ...757L...7L}. As the orbital trajectories contributing to this morphology intersect the plane in a very specific manner \citep{2002MNRAS.333..861S,2002MNRAS.337..578P}, empirical study of these orbits could potentially constrain the  gravitational potential along the inner Galactic plane, a matter of fundamental astrophysics that is difficult to investigate due to factors such as crowding and interstellar extinction.  We note that this feature can be probed over all six dimensions of kinematic phase space -- \cite{2013ApJ...776...76P} mapped the streaming motions of the split RC in both transverse directions, $\mu_{b}$ and $\mu_{l}$, with both showing strong signals. 

Though the discovery of the split-RC showed great promise to further understanding of Galactic dynamics, early efforts at precision characterisation of the double RC hit a wall due to the various sources of noise in the color-magnitude diagram. This is no surprise, as there were several uncertainties that needed to be cleared up, which we argue to be the parameters of the red giant branch bump (RGBB),  as well as the extinction.The RGBB is an excess in the red giant luminosity function that occurs as the hydrogen-burning shell nears the convective envelope \citep{1997MNRAS.285..593C}.  \citet{2010ApJ...721L..28N} showed that the double RC behaves in an erratic manner toward negative longitudes, which they argued could be due to degeneracies between the faint RC and the bright RGBB. \citet{2011ApJ...730..118N} and \citet{2013ApJ...766...77N} proceeded to characterise the RGBB of the Galactic bulge, and demonstrated that its properties were not consistent with predictions from canonical stellar evolution models and current understanding of the bulge stellar population. We have adapted that methodology to double-RC sightlines to measure the RGBB parameters toward the higher latitude fields investigated in this work. Meanwhile, the complex matter of the reddening toward the bulge was largely resolved by \citet{2012A&A...543A..13G} and  \citet{2013ApJ...769...88N}, at least for the sightlines relevant to this investigation. 

An impressive recent analysis is that of \citet{2013MNRAS.435.1874W}, who mapped the bulge RC in K-band using $VVV$ photometry \citep{2012A&A...537A.107S}. They mitigated the systematic errors mentioned above by independently measuring the reddening, and using the same RGBB parameters as  \citet{2013ApJ...769...88N}.  Their analysis benefits from broad coverage in both longitude and latitude, a rigorous treatment of errors, and an innovative non-parametric means to infer the geometrical distribution of bulge stars from their observed brightness and star count distribution functions.  They reported a best-fit value for the bar's angle of  $\alpha_{\rm{Bar}} = (27\pm 2)^{\circ}$, that the X-shape begins at a height of $\sim$400 pc from the plane, and that the Milky Way has a sharp X-shape. In contrast, it is estimated that only $\sim$33\% of bars in galaxies of Hubble type $ -2 \leq T \leq 1$ (de Vaucouleurs class S$0^{0}$ to Sa) contain an X-shaped structure \citep{2014arXiv1406.1418L}, respectively rising to $38 \pm 9$\% and $(62\pm9)$\% of early-type  $ T =2, 3$ (de Vaucouleurs class Sab, Sb) galaxies \citep{2013MNRAS.430.3489L}.

In this investigation, we empirically characterise the luminosity function of the stellar population toward these high-latitude sightlines, the RGBB parameters and the exponential slope of the underlying RGB luminosity function, which should be of use to future studies. We do this using $I$-band photometry, where systematic errors from stellar evolution and possible metallicity gradients are minimised \citep{2002MNRAS.337..332S}. We measure the photometric observables of the double-RC as a function of direction: the distance to the brighter RC, the brightness separation between the two RCs, the total surface density of RC stars on the sky,  and the fraction of stars in the brighter RC. We compare these results to predictions from two N-body models previously used to model dynamical data of the inner Galaxy.

\begin{figure*}
\begin{center}
\includegraphics[totalheight=0.4\textheight]{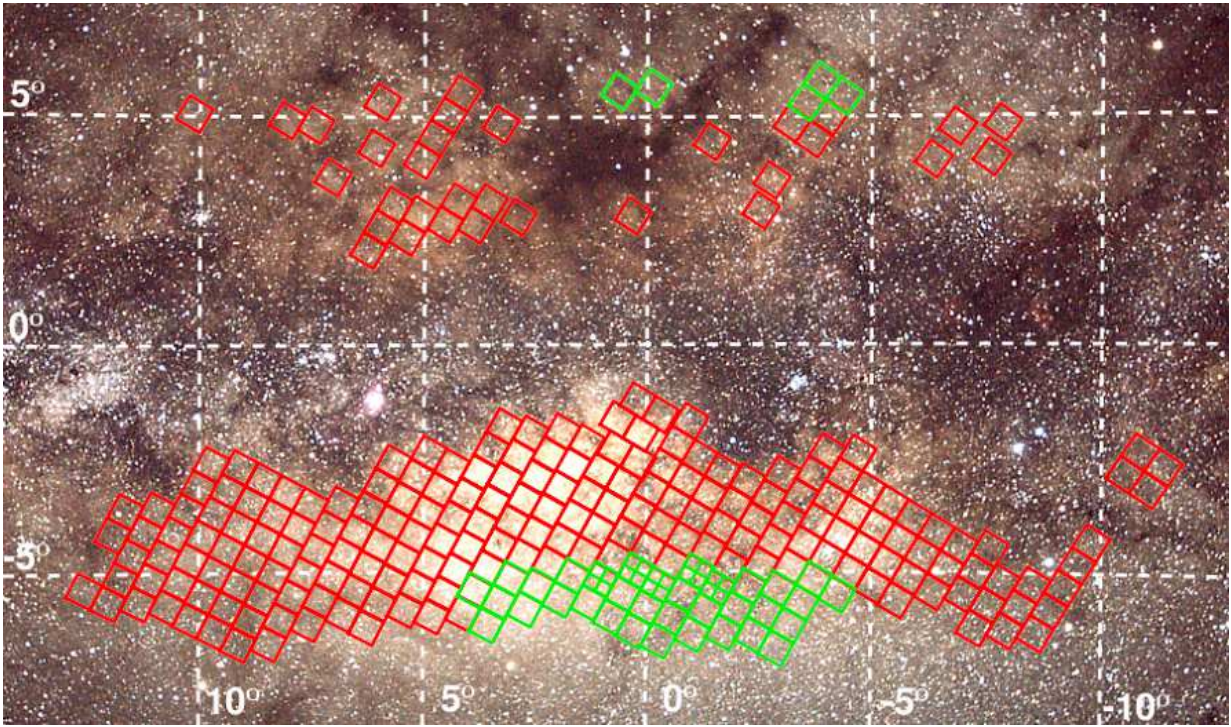}
\end{center}
\caption{\large Coverage of the OGLE-III Galactic bulge photometric survey used in this work, overplotted on an optical image of the same area. Galactic coordinate system shown.  The 54 OGLE-III calibration fields used in this work are shown as green squares, remaining fields are delineated by red squares.} 
\label{Fig:ogle3-blg-fields}
\end{figure*}

\section{Data}
\label{sec:Data}
OGLE-III observations were taken with the 1.3 meter Warsaw Telescope, located at the Las Campanas Observatory. The camera has eight 2048x4096 detectors, with a scale of approximately 0.26$\arcsec$/pixel yielding a combined field of view $0.6^{\circ}\times0.6^{\circ}$. We use observations from 39 of the 267 OGLE-III fields directed toward the Galactic bulge, where we select those for which the coordinates of the field centre satisfy  $(-5.00^{\circ} \leq  l_{\rm{central}} \leq 4.00^{\circ}, |b_{\rm{central}}| \geq 5.00^{\circ})$, keeping only those stars located on CCD chips (every OGLE-III field is made of 8 CCD chips) whose own centres satisfy that criteria. We deredden every star in those fields with the interpolated OGLE-III reddening and extinction maps of \citet{2013ApJ...769...88N}. Individual OGLE-III fields were selected as the investigative fields for our analysis as they were found to be a good compromise between the larger field sizes (and thus larger star counts) desirable for precise measurements, and the smaller field sizes desirable so as not to smooth over gradients. For five OGLE-III fields  (blg152, 159, 168, 176, and 186) for which coordinates of the field centres satisfy  $(-1.75^{\circ} \leq  l_{\rm{central}} \leq 1.25^{\circ}, 5.00^{\circ} \leq |b_{\rm{central}}| \leq 5.50^{\circ})$,  we split the OGLE-III field into four investigative fields (equivalent to two CCD chips), as the density of stars toward these sight lines is sufficiently high to allow a higher resolution in coordinate space, for a total of $34+20=54$ investigative fields, hereafter also referred to as calibration fields where appropriate. Our analysis is subsequently extended to 33 other OGLE-III fields toward larger longitudinal separations from the Galactic minor axis, for a total of 87 directions with measured parameters. 

The photometric coverage used in this work is shown in Figure \ref{Fig:ogle3-blg-fields}. More detailed descriptions of the instrumentation, photometric reductions and astrometric calibrations are available in  \citet{2003AcA....53..291U}, \citet{2008AcA....58...69U} and \citet{2011AcA....61...83S}. OGLE-III photometry and reddening maps are available for download from the OGLE webpage\footnote{http://ogle.astrouw.edu.pl/}.

\section{Measuring the Parameters of the Double Red Clump}
\subsection{Overview of Fitting}
\label{sec:RCmeasurements}
\begin{figure}
\begin{center}
\includegraphics[totalheight=0.18\textheight]{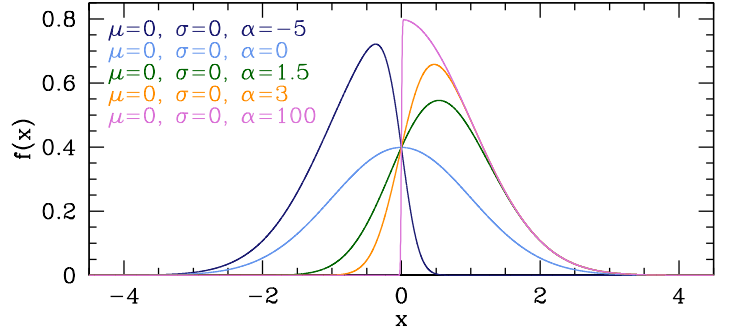}
\end{center}
\caption{\large Skew-Gaussian distributions with $(\mu,\sigma)=(0,1)$ plotted for demonstrative values of $\alpha$, including the standard-normal case $\alpha=0$, plotted in light blue. The $\alpha=100$ case is visually indistinguishable from a half-Gaussian. Note that the parameters $(\mu,\sigma)$ are only equal to $(\xi,\omega)$ in the case $\alpha=0$.} 
\label{Fig:SkewNormals}
\end{figure}

Modelling and interpreting the distribution of stars toward sightlines with a double RC is a non-trivial task due to the large number of features in the colour-magnitude diagram (CMD). There are two RCs, two RGBBs and two asymptotic giant branch bumps (AGBBs) mixed in with an underlying exponential-like luminosity function for the RGB and AGB stars, necessitating a diligent approach to avoid catastrophic degeneracies. All of these features are within our CMD selection box, which includes all stars redder than $(V-I) - (V-I)_{RC} >= -0.30$ and $|I-I_{RC}| \leq 1.50$, where the mean values  $ (V-I)_{RC} ,I_{RC} $ are as determined by \citet{2013ApJ...769...88N}.    We first attempted to model the luminosity function of the red giant branch using the same formalism as  \citet{2013ApJ...769...88N}, \citet{2013ApJ...776...76P}, and \citet{2013MNRAS.435.1874W} but we found that it was too costly to not fit for the skewness of the RCs. We have thus replaced the Gaussian distributions for the RCs with skew-Gaussian distributions, for which we fix the skew to be equal and opposite. Whereas a standard Gaussian distribution function is given by:
\begin{equation}
\phi(x) =  \frac{1}{\sqrt{2\pi}} \exp{ \biggl[ \frac{-x^2}{2}  \biggl]},
\label{EQ:equationforGaussian}
\end{equation}
a skew-Gaussian is given by:
\begin{equation}
f(x) =  2 \phi(x)  \int_{-\infty}^{\alpha_{\rm{Skew}} x} \phi(t) \, dt,
\label{EQ:equationforSkewGaussian}
\end{equation}
where location and scale can be added in using the transformation:
\begin{equation}
x \rightarrow \frac{x-\xi}{\omega},
\label{EQ:equationforSkewGaussianLocScale}
\end{equation}
leading us to adopt the notation $f(\omega,\xi,\alpha_{\rm{Skew}})$ for the skew-Gaussian distributions, with parameters:
\begin{equation}
\delta = \frac{\alpha_{\rm{Skew}}} {\sqrt{1+{\alpha_{\rm{Skew}}}^2  } }
\end{equation}
\begin{equation}
\rm{Mean}\,=I_{RC}=\,\xi+\omega\delta\sqrt{\frac{2}{\pi}}
\end{equation}
\begin{equation}
\rm{Variance}\,= \sigma_{RC}^2=\, \omega^2 \biggl( 1 - \frac{2\delta^2}{\pi}    \biggl)
\end{equation}
\begin{equation}
\rm{Skewness}\,=\, \frac{4-\pi}{2}  \frac{ \biggl( \delta\sqrt{2/\pi}   \biggl)^3   }{    \biggl(  1-2\delta^2/\pi      \biggl)^{3/2}}
\end{equation}
In the case $\alpha_{\rm{Skew}}=0$, the skew-Gaussian distribution reduces to a standard Gaussian distribution, with parameters $(\mu,\sigma)=(\xi,\omega)$. Conversely, as $\alpha_{\rm{Skew}}$ approaches $\pm \infty$, the skew-Gaussian distribution converges to a half-Gaussian distribution, with skew approaching $\gamma_{1}= \pm  \sqrt{2}(4-\pi)/(\pi-2)^{2} \approx \pm 0.995$.  A set of demonstrative skew-Gaussian distributions are plotted in Figure \ref{Fig:SkewNormals}, including the $\alpha_{\rm{Skew}}=1.5$ case assumed by our analysis. We note that given the demonstrative values $\mu=0$ and $\sigma=1$ with $\alpha_{\rm{Skew}}=1.5$ will yield a mode of $x=0.543$, a median of $x=0.624$, a mean of $x=0.664$, a standard deviation of 0.748 and a skewness of 0.300. 

We parametrize the luminosity function as follows:
\begin{equation}
\begin{split}
N(m) =  \sum_{i=1}^{i=2} \biggl\{  A_{i}\exp\biggl[B_{i}(I-I_{RC_{i}})\biggl] +  N_{RC, i}f(\omega_{i},\xi_{i},\alpha_{Skew,i}) + \nonumber \\
 \frac{N_{RGBB, i}}{\sqrt{2\pi}\sigma_{RGBB}}\exp \biggl[{-\frac{(I-I_{RGBB, i})^2}{2\sigma_{RGBB}^2}}\biggl]  \\ 
 + \frac{N_{AGBB, i}}{\sqrt{2\pi}\sigma_{AGBB}}\exp \biggl[{-\frac{(I-I_{AGBB, i})^2}{2\sigma_{AGBB}^2}}\biggl] \biggl\}, \\ 
\end{split}
\label{EQ:Exponential}
\end{equation}
where the parameters are as defined by \citet{2013ApJ...769...88N}.

We reduce the number of free parameters by imposing the following constraints:
\begin{enumerate}
\item The two RGBs have the same exponential slope parameter $B$, and the same ratio $EW_{RC} = N_{RC,i}/A_{i}$ (2 constraints).  
 \item The integral of the number of stars in the parametrized luminosity function is made equal to the number of stars observed in the color-magnitude selection box (1 constraint).
 \item Both AGBBs are fixed to have 2.8\% the number of stars of their respective RCs, to be 1.07 mag brighter than their respective RCs, and to have the same magnitude dispersion as their respective RCs (6 constraints), as in \citet{2013ApJ...769...88N}. 
 \item The skewness of the two RCs is equal and opposite, with the skewness of the brighter RC being strictly negative, i.e. a long tail toward brighter magnitudes (1 constraint). This is explored in Section \ref{sec:SkewAssumptions}. 
  \end{enumerate}
There are 14 free parameters remaining following these  constraints, which is still too large. 

\subsection{Further Constraints on the Red Giant Luminosity Function}
\label{subsec:RelativeDispersions}
We constrain the exponential slope of the RG luminosity function $B$, the mean value of the skewness parameter $\alpha_{\rm{Skew}}$, the ratio of the magnitude dispersion of the faint RC to that of the bright RC $\sigma_{RC,2}/\sigma_{RC,1}$, and the RGBB parameters ${\Delta}I^{RC}_{RGBB} =  I_{RGBB}-I_{RC} $, ${f}^{RC}_{RGBB} =  N_{RGBB}/N_{RC}$, and $ {\Delta\sigma}^2 = \sigma_{RGBB}^2-\sigma_{RC}^2$. The latter five parameters were iterated in a grid search, whereas $B$ was allowed to float. 

The best-fit values found for the fainter RC-RGBB pair are:
\begin{equation}
 {\Delta}I_{RGBB}^{RC} = 0.610,  \vspace{-0.3cm}
\end{equation}
\begin{equation}
 f_{RGBB}^{RC} = 0.160, \vspace{-0.3cm}
\end{equation}
\begin{equation}
{\Delta\sigma}^2 =  0.0050, \vspace{-0.3cm} 
\end{equation}
\begin{equation}
\sigma_{RC,2} / \sigma_{RC,1} = 0.80, \vspace{-0.3cm}
\end{equation}
\begin{equation}
 \alpha_{\rm{Skew}}  = +1.50, \vspace{-0.1cm}
\end{equation}
As stated before the two RCs have equal and opposite skew. For the brighter RC-RGBB pair, which is closer to the plane at fixed latitude and thus has a higher metallicity (${\Delta}$[M/H]$\approx$0.06 dex or ${\Delta}$[Fe/H]$\approx$0.08 dex, \citealt{2013MNRAS.430..836N}), we raise ${\Delta}I_{RGBB}^{RC}$ by 0.0535 mag and increase $ f_{RGBB}^{RC} $ by 0.007, as per the empirical calibrations of \citet{2013ApJ...769...88N}. We also fix the exponential parameter
\begin{equation}
B=0.578,
\end{equation}
which is both the mean and median value found, independently of the values of the other parameters. The combination of the six constraints on the RGBB, the constraint on $B$, and on the relative magnitude of the dispersion of the two RCs, leaves us with just five free parameters per field: $\sigma_{RC,1}$, $I_{RC,1}$, $I_{RC,2}$, $N_{RC,1}$, and $N_{RC,2}$, though what we report is the brightness to the brighter RC, the brightness separation between the two RCs, the total surface density of RC stars on the sky,  and the fraction of stars in the brighter RC. 

Some of these constraints can be interpreted in a straightforward manner. The values of our RGBB parameters are consistent with a difference in the metallicity ${\Delta}$[M/H]$\approx$0.22 dex (or ${\Delta}$[Fe/H]$\approx$0.27 dex)  between $(l,b)=(0^{\circ},-2^{\circ})$ (measured by \citealt{2013ApJ...769...88N}) and the sightlines investigated here, for which $(|l|  \lesssim 3.5^{\circ}   , 5^{\circ} \lesssim  |b| \lesssim 7^{\circ})$. The dispersion for the fainter RC will be smaller than that of the brighter RC if the brightness dispersion is dominated by geometric dispersion, given that the fainter RC is further away. A distance modulus difference of ${\delta}{\mu}=0.50$ mag yields $\sigma_{\mu,2}/\sigma_{\mu,1} \approx 10^{-0.1}=0.79$. As this is consistent with the value found here of $ \sigma_{RC,2} / \sigma_{RC,1} = 0.80 \pm 0.02$, it is likely that the observed magnitude dispersion of RC stars is dominated by geometric dispersion, and not by stellar evolution, photometric noise, or residual differential reddening. This constraint is explored in Section \ref{sec:SkewAssumptions}. 

\begin{figure}
\begin{center}
\includegraphics[totalheight=0.33\textheight]{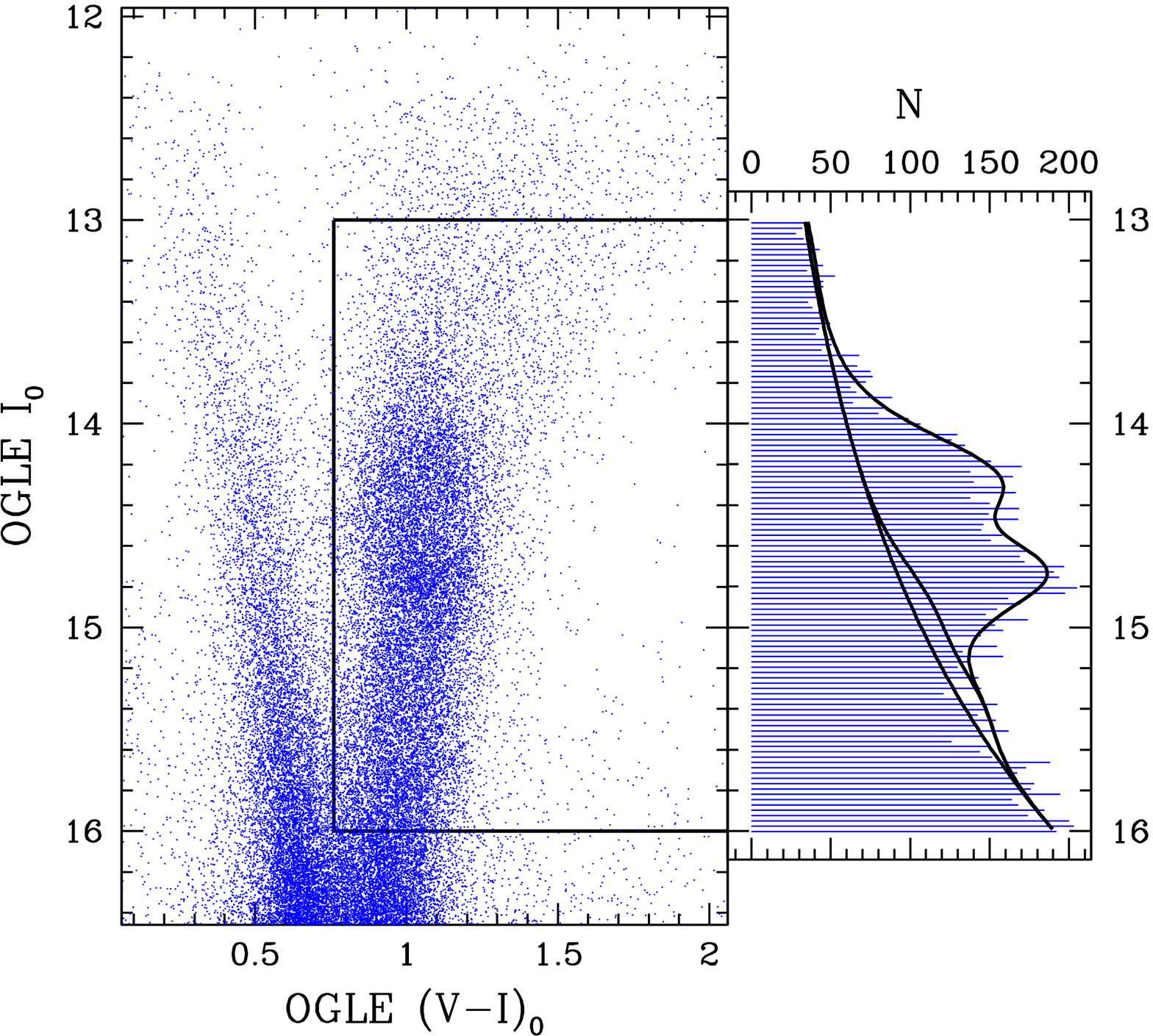}
\end{center}
\caption{\large LEFT: Dereddened colour-magnitude diagram for the field BLG154, toward $(l,b)=(-1.04^{\circ},-6.38^{\circ})$. The colour-magnitude selection box is delineated by the thick, black rectangle.  The bright red clump has a dereddened peak magnitude of $I_{RC,1}=14.24 \pm 0.01$, and the magnitude difference between the two red clumps is ${\Delta}I=0.52\pm0.01$. The fraction of stars in the faint red clump, $N_{\rm{Faint}}/N_{\rm{Total}}$, is $(43\pm2)$\%.   RIGHT: Luminosity function for red giant and red clump stars.  } 
\label{Fig:MarkovPresentationPlot}
\end{figure}

\begin{figure}
\begin{center}
\includegraphics[totalheight=0.33\textheight]{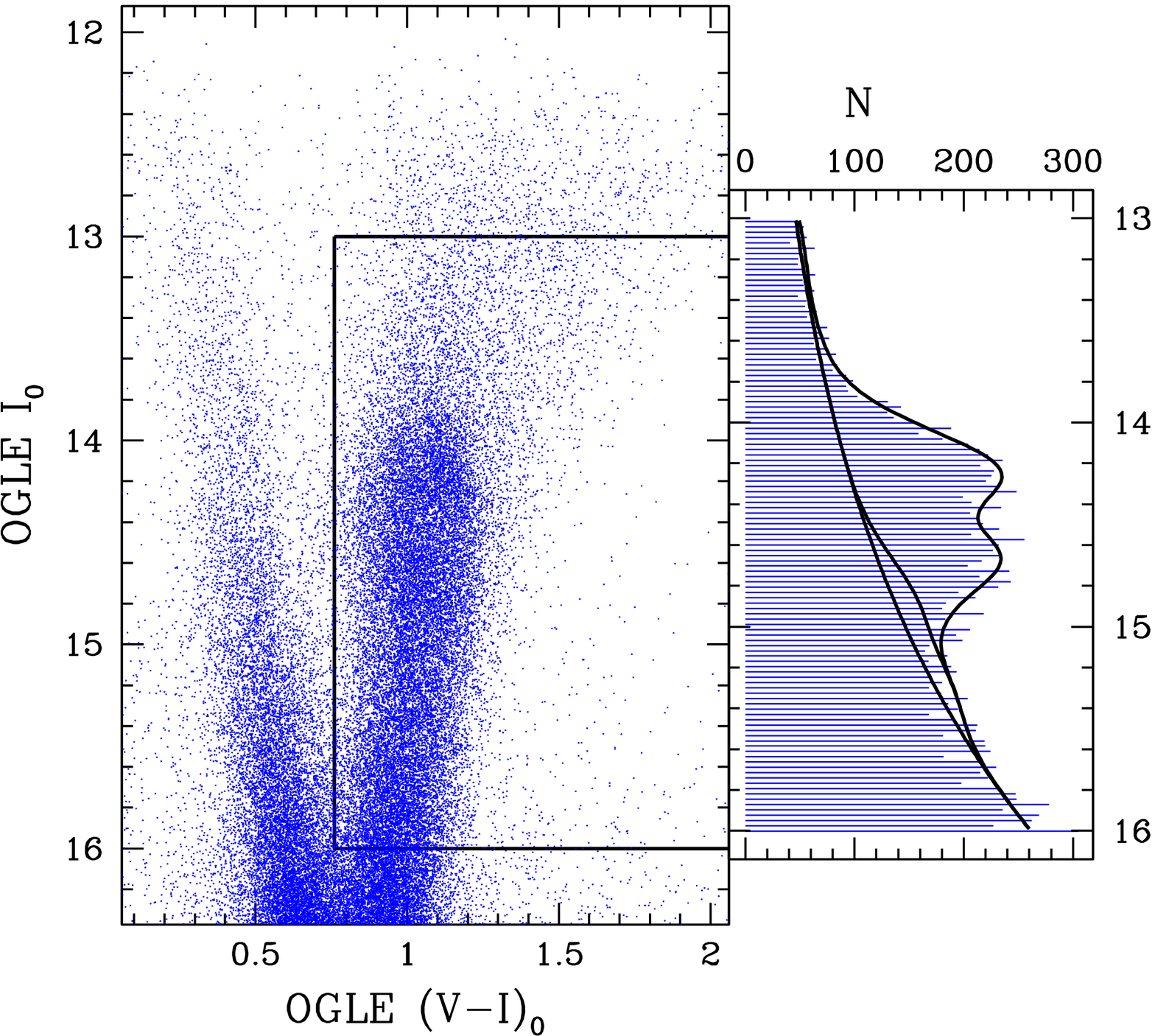}
\end{center}
\caption{\large LEFT: Dereddened colour-magnitude diagram for the field BLG170, toward $(l,b)=(0.27^{\circ},-6.31^{\circ})$. The colour-magnitude selection box is delineated by the thick, black rectangle.  The bright red clump has a dereddened peak magnitude of $I_{RC,1}=14.19 \pm 0.01$, and the magnitude difference between the two red clumps is ${\Delta}I=0.47\pm0.01$. The fraction of stars in the faint red clump, $N_{\rm{Faint}}/N_{\rm{Total}}$, is $(36\pm2)$\%.   RIGHT: Luminosity function for red giant and red clump stars.  } 
\label{Fig:MarkovPresentationPlot2}
\end{figure}

\begin{figure}
\begin{center}
\includegraphics[totalheight=0.33\textheight]{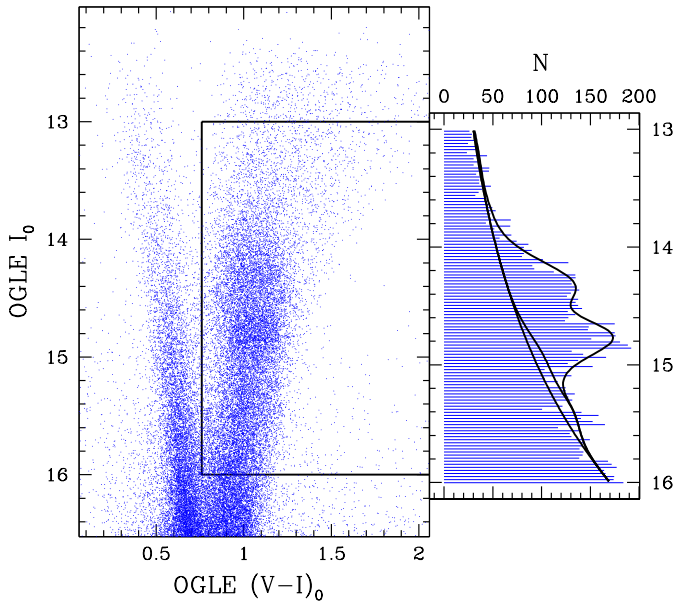}
\end{center}
\caption{\large LEFT: Dereddened colour-magnitude diagram for the field BLG128, toward $(l,b)=(-2.90^{\circ},-6.68^{\circ})$. The colour-magnitude selection box is delineated by the thick, black rectangle.  The bright red clump has a dereddened peak magnitude of $I_{RC,1}=14.29 \pm 0.01$, and the magnitude difference between the two red clumps is ${\Delta}I=0.50\pm0.01$. The fraction of stars in the faint red clump, $N_{\rm{Faint}}/N_{\rm{Total}}$, is $(49\pm2)$\%.   RIGHT: Luminosity function for red giant and red clump stars.  } 
\label{Fig:MarkovPresentationPlot3}
\end{figure}

The clarity resulting from improved reddening maps and a more robust parameterization can be discerned in Figure \ref{Fig:MarkovPresentationPlot}, where we show the dereddened CMD for the OGLE-III field BLG154. The field contains $4006\pm120$ RC stars, yielding a convincing signal for not just the split-RC, but many of its features. We also plot the CMDs for BLG170 and BLG128 in Figure \ref{Fig:MarkovPresentationPlot2} and \ref{Fig:MarkovPresentationPlot3}.

\section{N-Body Models}
\label{Sec:Nbody}
In order to facilitate interpretation of our data, we have convolved a scaled-solar, $t=11$ Gyr, [M/H]$=+0.06$ theoretical RG+RC luminosity function from the BaSTI stellar database   \citep{2004ApJ...612..168P,2007AJ....133..468C}\footnote{BaSTI models can be found at the following URL:  http://basti.oa-teramo.inaf.it/index.html} with two different N-body models, and then we applied our algorithms as described in Section \ref{sec:RCmeasurements} to see if we can recover similar structural parameters. Our luminosity function assumes a Salpeter initial mass function (IMF) $dn/dm \propto m^{-2.35}$ \citep{1955ApJ...121..161S}. In practice, the choice of IMF is predicted to have a negligible impact on RG luminosity functions, as evolution along the RG branch is fast \citep{2006ApJ...641.1102B}.

\begin{figure}
\begin{center}
\includegraphics[totalheight=0.35\textheight]{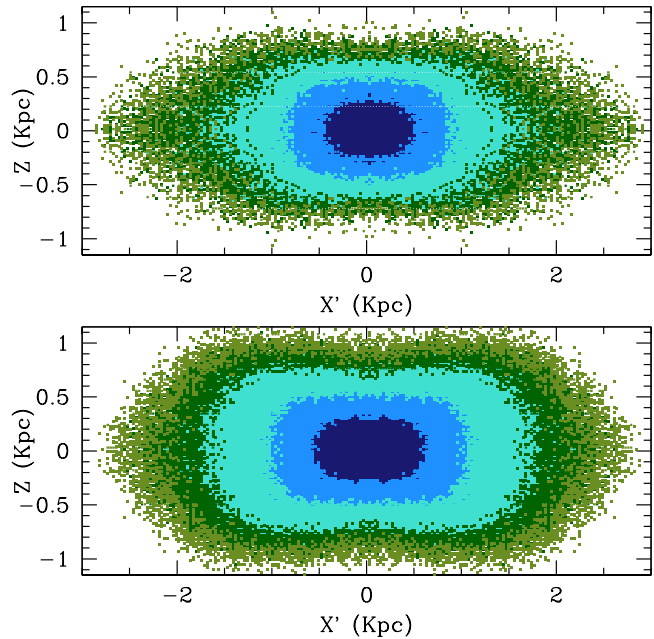}
\end{center}
\caption{\large  Density contours for the two N-body models used in this work,  that of \citet{2012ApJ...756...22N}  is shown in the top panel and that of  \citet{2010ApJ...720L..72S} is shown in the bottom-panel. Both models are viewed side-on, such the major axis of the bar is perpendicular to the line of sight. The vertical axis is denoted  and the planar axis parallel to the bar's major axis is denoted X'. Colours denote the pixels containing 25\%, 50\%, 75\%, 85\% and 91\% of the stars in these models within the coordinate range $\sqrt{(X^2 + Y^2)} \leq 3.0\,\rm{Kpc}, |Z| \leq 1.50\,\rm{Kpc}$, i.e. 25\% of the stars in this coordinate range lie within the pixels drawn in dark blue points. } 
\label{Fig:NbodyDensityContours4paper}
\end{figure}

\begin{figure}
\begin{center}
\includegraphics[totalheight=0.35\textheight]{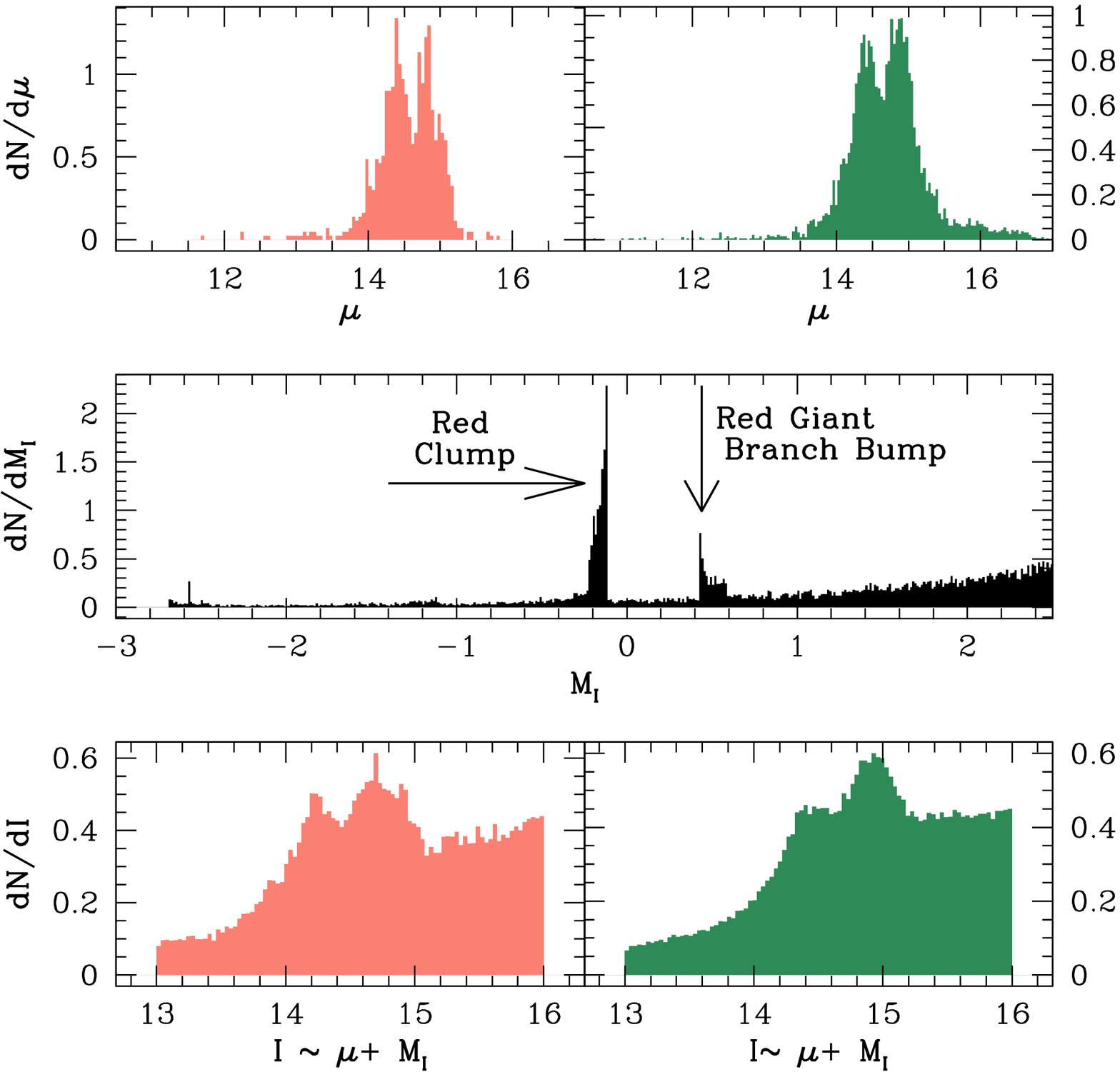}
\end{center}
\caption{\large TOP: Predicted distance modulus distribution functions  toward $(l,b)=(-1.00^{\circ},-6.00^{\circ})$ for the models of \citet{2012ApJ...756...22N} (Salmon) and  \citet{2010ApJ...720L..72S} (seagreen).  MIDDLE:  Scaled-solar, $t=11$ Gyr, [M/H]$=+0.06$ theoretical RG+RC luminosity function from the BaSTI stellar database   \citep{2004ApJ...612..168P,2007AJ....133..468C} with which we convolve the N-body models. BOTTOM: Resultant apparent magnitude distribution function, where we have also added 0.05 mag of Gaussian noise to simulate the effects of secondary noise sources. } 
\label{Fig:NbodyLFHist}
\end{figure}

The two N-body models studied in this work have been selected as they have already been used in the bulge-related literature that has furthered understanding of Galactic dynamics.  

We use the same N-body model as \citet{2012ApJ...756...22N}, which consists of two live components, namely the halo and the disk, and is fully self-consistent.  It was originally studied and developed by \citet{2003MNRAS.341.1179A} and described in those works.  \citet{2012ApJ...756...22N} showed that the model includes a bimodal density peak along the line of sight, and approximately reproduces the sign and amplitude of the radial velocity differences between the two density peaks along the line of sight. At the time of the adopted snapshot, the corotation radius of the bar is $\sim$6.3 Kpc and the circular velocity at that radius is $\sim$210 km/s. 

We also use the same N-body model as \citet{2010ApJ...720L..72S}, which was also studied by \citet{2012ApJ...757L...7L} , by  \citet{2014ApJ...785L..17L} and by \citet{2014IAUS..298..201S}. The model of  \citet{2010ApJ...720L..72S} has been shown to match the directional-dependence of the first and second moments of the radial velocity distributions from the  the BRAVA survey \citep{2012AJ....143...57K}, to have an X-shaped bulge, and to have a radial velocity distribution function that is skew-symmetric for sightlines close to the plane.  At the time of the adopted snapshot, the corotation radius of the bar is $\sim$4.5 Kpc and the circular velocity at that radius is $\sim$180 km/s. A side-on view of the two models is shown in Figure \ref{Fig:NbodyDensityContours4paper}.

Both models evolved from initial conditions with dynamically cold disks, such that Toomre's Q $\approx 1.2$. 
In both cases, we adopt the same scale factors as assumed by those works in order for our comparisons to be consistent.  We apply the 8-fold symmetry to both models to increase the effective star counts, and thus we only ``observe" negative latitudes. We assume a distance to the Galactic Centre of $R_{\rm{GC}}=8.36$ Kpc from \citet{2014arXiv1403.5266C}, who quoted an error of $0.11$ Kpc. That measurement is based on a combined Bayesian analysis of the statistical parallax to the Milky Way's old nuclear star cluster as well as the orbits of stars near Sgr A* from \citet{2009ApJ...692.1075G}. This distance is consistent with the recent estimate of $R_{\rm{GC}} = (8.27 \pm 0.29)$ Kpc \citep{2012MNRAS.427..274S}, which is derived from solar neighbourhood kinematic data based on the assumption that the Milky Way disk is nearly axisymmetric. 

\begin{table*}
\caption{\large Comparison of distance scales between the data and the models. } 
\large
\centering 
\begin{tabular}{|cccc|} 
\hline\hline\hline 
Category &  $I_{RC,\rm{Faint}}- I_{RC,\rm{Bright}}$ &  $d\log{\Sigma_{RC}}/d|b|$ & Scale Height \\
 &  toward $(l,b)=(0^{\circ},-5.5^{\circ})$ & deg$^{-1}$  & (Kpc) \\  
 \hline\hline
Data &  0.48 & $-0$.246 & N/A \\
N2012 model, $\alpha_{\rm{Bar}}=30^{\circ}$ &  0.49 & $-$0.284 & 0.254  \\
S2010 model, $\alpha_{\rm{Bar}}=30^{\circ}$ &  0.50 & $-$0.206 & 0.366  \\ 
\hline\hline
\end{tabular}
\label{table:DistanceScales} 
\end{table*}

\begin{table*}
\caption{\large Predicted skewness values and magnitude dispersions from the N-body models. } 
\large
\centering 
\begin{tabular}{|cc|ccc|ccc|} 
\hline\hline\hline 
$l$ & $b$ & Skew$_{1}$ & Skew$_{2}$ &  $\sigma_{RC,2} / \sigma_{RC,1}$ & Skew$_{1}$ & Skew$_{2}$ &  $\sigma_{RC,2} / \sigma_{RC,1}$  \\
 - & - &  N2012 & N2012 & N2012 &  S2010 & S2010 & S2010  \\
 \hline\hline
-3.00 & -6.00 & -0.91 & 0.66 & 0.78  & -0.99 & 0.47 & 0.85    \\ 
-2.00 & -6.00 & -0.90 & 0.55 & 0.78  & -0.92 & 0.48 & 0.86   \\ 
-1.00 & -6.00 & -0.81 & 0.52 & 0.79   & -0.92 & 0.50 & 0.88    \\ 
0.00 & -6.00 & -0.91 & 0.68 & 0.76  & -0.93 & 0.45 & 0.89     \\ 
1.00 & -6.00 & -0.92 & 0.49 & 0.83  & -0.91 & 0.51 & 0.96    \\ 
2.00 & -6.00 & -0.90 & 0.73 & 0.82    & -0.98 & 0.52 & 0.98    \\ 
\hline\hline
\end{tabular}
\label{table:SkewTest} 
\end{table*}

The approach used here can best be discerned by inspecting Figure \ref{Fig:NbodyLFHist}. We show a distance modulus distribution function for a sightline as predicted by the N-body models (top two panels), the absolute magnitude distribution function assumed (middle panel), and the final luminosity function that is the convolution of the intrinsic distance distribution and absolute magnitude distribution, with 0.05 mag of Gaussian noise added to simulate the effects of photometric errors, binaries, metallicity dispersion, and residual differential extinction (bottom panels).  For these fits, we fix the AGBB and AGBB parameters to be the same as they are in the intrinsic luminosity function, the exponential slope parameter to $B=0.60$ \citep{2014MNRAS.442.2075N}. We fix $\sigma_{RC,2} / \sigma_{RC,1} = 0.80$, $\alpha_{\rm{Skew}}=+1.50$ for the faint RC, and $\alpha_{\rm{Skew}}=-1.50$ for the bright RC, as we did for the observations.  Though it would be interesting to independently derive these parameters in the model, the number of stars in the models by   \citet{2012ApJ...756...22N} and \citet{2010ApJ...720L..72S} with $(|l|  \lesssim 3.5^{\circ}   , 5^{\circ} \lesssim  |b| \lesssim 7^{\circ})$ are $\sim$14,000 and $\sim$78,000 respectively, compared to $\sim$220,000 RC+RGBB stars in the observations. 

\subsection{On the Selected Length Scale for the Models}
The normalisation for distances (and velocities) in N-body models is arbitrary, and will depend on factors such as the assumed distance to the Galactic Centre, and the viewing angle to the Galactic bar. In particular, it will depend on which physical parameter is adopted to set the scale, as different parameters should yield different scale factors, since no N-body model is expected to be a perfectly scaled-replica of the Milky Way. \citet{2012ApJ...756...22N} scaled their N-body model by matching to the surface brightness maps from COBE \citep{1995ApJ...445..716D}, and \citet{2010ApJ...720L..72S}  calibrated their length scales by matching their models to the radial velocity profile with direction measured by the BRAVA survey \citep{2008ApJ...688.1060H,2012AJ....143...57K}.

In Table \ref{table:DistanceScales}, we show the relative distances achieved for two other observables we could have used to set the scale factors: the brightness separation between the bright and faint RC toward $(l,b)=(0^{\circ},-5.5^{\circ})$, and the observed latitudinal scale length of the surface density of RC stars, $d\log{\Sigma_{RC}}/d|b|$, toward $(-3.5^{\circ}<=l<=2.5^{\circ}, 5.0^{\circ}<=|b|<=7.0^{\circ})$. Impressively, the predicted brightness separation between the two peaks is a great match to the data for both models when the models oriented at an angle of $\alpha_{\rm{Bar}}=30^{\circ}$, which is consistent with the observational constraints \citep{2013MNRAS.434..595C,2013MNRAS.435.1874W}.

We note that the observed value of $d\log{\Sigma_{RC}}/d|b|$ has a first-order mapping onto the exponential scale height, given by:
\begin{equation}
\frac{dZ}{d\log_{e} {\Sigma}}    =  \frac{\pi}{180}     \frac{R_{\rm{GC}}  }{\log_{e}{(10)}     }     \biggl( \frac {d \log \Sigma }{d|b|} \biggl)^{-1}\approx 0.26 \, \rm{Kpc}.
\label{EQ:ScaleHeight}
\end{equation} 
This estimated scale height is marginally smaller than that of 0.27 Kpc reported by \citet{2013MNRAS.434..595C}. Comparison to the analysis of \citet{2013MNRAS.435.1874W} are more difficult as their scale height is position-dependent  (though they determine a global-average scale height of 0.18 Kpc in their abstract). We restrict the comparison to their trend for the minor axis (the blue curve of their Figure 15), from which we estimate a scale height of $\sim$0.33 Kpc. This is $\sim$27\% greater than the naive estimate of Equation \ref{EQ:ScaleHeight}.

However, if we repeat this calculation with one of our N-body models \citep{2012ApJ...756...22N}, given an average latitude of $|b|=6^{\circ}$, $d\log{\Sigma_{RC}}/d|b| = -0.284$ (see the second row of Table \ref{table:DistanceScales}), we obtain a scale height of $\sim$223 pc. That answer may appear plausible, but it is not the true scale height. The actual scale height in the model of \citet{2012ApJ...756...22N}  for  $0.70 <= Z/\rm{Kpc} <= 1.00$ and $\sqrt{(X^2 + Y^2)} \leq 3.0\,\rm{Kpc}$ is 254 pc, a $\sim$14\% offset relative to the naive estimate.  This discrepancy arises due to at least two reasons. First, the number density along the line of sight is bimodal, with the two peaks being at different heights for a given latitude. Second, within the models and plausibly within the Galaxy, scale height is a sensitive, non-linear function of Galactocentric radius, as such there will be some fuzziness when reporting any specific scale height.


\subsection{On The Assumption of Equal and Opposite Skew and a Fixed Magnitude Dispersion Ratio}
\label{sec:SkewAssumptions}
The availability of N-body models yields a means of verifying the reliability of three  assumptions imposed in Section \ref{subsec:RelativeDispersions},  that the two skews of the two RCs are equal and opposite, and that the ratio of their magnitude dispersions is fixed, such that $\sigma_{RC,2} / \sigma_{RC,1} = 0.80$ for all of our sightlines. It is non-trivial to precisely investigate this in the data as these parameters are strongly degenerate, both with each other but also with other parameters such as $f_{RGBB}^{RC}$ and $ {\Delta}I_{RGBB}^{RC}$. 

We model the skew by taking the distance modulus distribution functions for the N-body models, under the assumption that $\alpha_{\rm{Bar}}=30^{\circ}$, and convolve them with Gaussian noise of 0.05 mag. We partition the magnitudes into the ranges $13.61 <   \mu_{1} \leq 14.61$ and   $14.61 <   \mu_{1} \leq 15.41$, where the dividing point of $\mu=14.61$ is from our assumed distance to the Galactic centre of $R_{\rm{GC}}=8.36$ Kpc \citep{2014arXiv1403.5266C}, and is thus consistent with the position of the bifurcation, see the top panels of Figure \ref{Fig:NbodyLFHist}. The results are listed in Table \ref{table:SkewTest}. We also experiment with other choices of limits and magnitude dispersion convolutions. What remains consistent are that the skews are both opposite in sign and large, $0.50 \leq \, \rm{Skew}\, < 1.00$ (corresponding to  $2 \lesssim \, \alpha_{\rm{Skew}}   \, <  \infty $), and that  $0.60 \lesssim \sigma_{RC,2} / \sigma_{RC,1} \lesssim 1.05$. The model of \citet{2010ApJ...720L..72S} is better-fit by larger values of $ \sigma_{RC,2} / \sigma_{RC,1}$ than that of \citet{2012ApJ...756...22N}. That is likely due to the larger scale height of the bar in the model of \citet{2010ApJ...720L..72S}  (see Figure \ref{Fig:NbodyDensityContours4paper}), which leads to a more populous faint RC relative to the bright RC. 

Thus, the N-body model predictions are slightly offset from our best-fit values used in Section \ref{subsec:RelativeDispersions}, and even further offset from the assumption of two Gaussians or even two Gaussians of equal magnitude dispersions ubiquitously assumed in the literature prior to this investigation. However, all of these assumptions are less offset from each other than they would be from the the severe, unphysical, degeneracy-induced variations that would be ``measured" from one sightline to the next if no constraints were imposed. As the discrepancy may come from factors other than the geometry of the bar, for example undiagnosed sources of noise in the colour-magnitude diagram, or the greater incidence of disk contamination seen in the data relative to the models (see Section \ref{sec:EWRC}), we opt to leave our assumptions for these two parameters as they are. The outstanding uncertainty in these parameters suggests that measurements of bulk properties of the combined RCs such as $EW_{RC}$ and $\Sigma_{RC}$ are likely more reliable than those of relative properties between the two RCs, such as  $N_{\rm{Faint}}/N_{\rm{Total}}$.

\section{Photometric Analysis of the Double Red Clump: Results}
\label{sec:Results}

Our key findings concern the brightness of the two RCs, the total surface density of RC stars deg$^{-2}$, $\Sigma_{RC}$, the fraction of RC stars in the faint RC, $N_{\rm{Faint}}/N_{\rm{Total}}$, and the ratio of RC stars to the number density of RG stars at the luminosity of the RC, $EW_{RC}$. We discuss each of these separately. Our results for our 54 calibration fields are enumerated in Table 3, with the results of the analysis extended to 33 other fields and Table 4. The error values are 1-$\sigma$ errors estimated via Markov Chain Monte Carlo, they are purely statistical and do not include the effects of systematic errors. 

The parameters toward the field blg159A are clearly erroneous as they differ remarkably from those of neighbouring fields, which may be a product of unresolved differential reddening toward that field. We have verified that this field does not impact our selection of the best-fit RGBB parameters, or any of the other parameters and conclusions discussed later in this work.

\begin{table*}
\normalsize
\caption{\large Parameters of split red clumps toward the 54 calibration fields, listed in order of increasing Galactic longitude. Listed are the OGLE-III field names, the field centres in Galactic coordinates, the dereddened apparent magnitude of the brighter red clump $I_{RC,1}$, the brightness difference between the two red clumps  (${\Delta}I$), the fraction of red clump stars in the faint red clump ($N_{\rm{Faint}}/N_{\rm{Total}}$), the total number of red clump stars $N_{RC}$, the surface density of red clump stars per deg$^{2}$, the number density of red giant stars at the luminosity of the red clump $EW_{RC}$, and the magnitude dispersion of the brighter red clump, $\Sigma_{RC}$. Some OGLE-III fields are split into components \{A,B,C,D\}.  } 
\centering 
\begin{tabular}{llllllllll}
\hline\hline\hline 
Field & $l$ & $b$ & $I_{RC,1}$ & ${\Delta}I$ & $N_{\rm{Faint}}/N_{\rm{Total}}$ & $N_{RC}$ & $\Sigma_{RC}$  & $EW_{RC}$ & $\sigma_{RC,1}$ \\
 \hline\hline
blg363 & -4.37 & 5.43 & 14.36$\pm$0.02 &  0.38$\pm$0.02 & 0.54$\pm$0.04 &  7130$\pm$ 167 & 20286$\pm$  475  & 1.08$\pm$0.04   & 0.23$\pm$0.01 \\
blg120 & -4.25 & -5.50 & 14.35$\pm$0.01 &  0.39$\pm$0.01 & 0.50$\pm$0.02 &  6968$\pm$ 138 & 19504$\pm$  386  & 1.10$\pm$0.03   & 0.20$\pm$0.01 \\ 
blg360 & -3.89 & 5.78 & 14.32$\pm$0.01 &  0.44$\pm$0.01 & 0.57$\pm$0.02 &  6236$\pm$ 142 & 17306$\pm$  394  & 1.07$\pm$0.03   & 0.21$\pm$0.01 \\ 
blg125 & -3.75 & -5.19 & 14.36$\pm$0.01 &  0.38$\pm$0.01 & 0.51$\pm$0.02 &  8855$\pm$ 157 & 24156$\pm$  428  & 1.20$\pm$0.03   & 0.20$\pm$0.01 \\ 
blg359 & -3.56 & 5.31 & 14.43$\pm$0.02 &  0.39$\pm$0.01 & 0.49$\pm$0.03 &  8463$\pm$ 165 & 23427$\pm$  458  & 1.18$\pm$0.03   & 0.22$\pm$0.01 \\ 
blg126 & -3.46 & -5.69 & 14.34$\pm$0.01 &  0.42$\pm$0.01 & 0.55$\pm$0.02 &  6526$\pm$ 137 & 17822$\pm$  374  & 1.08$\pm$0.03   & 0.20$\pm$0.01 \\ 
blg127 & -3.18 & -6.18 & 14.25$\pm$0.01 &  0.47$\pm$0.01 & 0.55$\pm$0.02 &  4590$\pm$ 121 & 13187$\pm$  346  & 1.00$\pm$0.04   & 0.20$\pm$0.01 \\ 
blg135 & -2.95 & -5.38 & 14.30$\pm$0.01 &  0.43$\pm$0.01 & 0.55$\pm$0.01 &  8068$\pm$ 155 & 21929$\pm$  420  & 1.14$\pm$0.03   & 0.20$\pm$0.01 \\ 
blg128 & -2.90 & -6.68 & 14.29$\pm$0.01 &  0.50$\pm$0.01 & 0.49$\pm$0.02 &  3252$\pm$ 107 &  9704$\pm$  319  & 0.90$\pm$0.04   & 0.20$\pm$0.01 \\ 
blg136 & -2.67 & -5.89 & 14.28$\pm$0.02 &  0.46$\pm$0.01 & 0.51$\pm$0.02 &  6207$\pm$ 148 & 16995$\pm$  406  & 1.09$\pm$0.04   & 0.23$\pm$0.01 \\ 
blg137 & -2.39 & -6.39 & 14.25$\pm$0.01 &  0.49$\pm$0.01 & 0.51$\pm$0.01 &  4269$\pm$ 118 & 11815$\pm$  326  & 0.96$\pm$0.03   & 0.20$\pm$0.01 \\ 
blg144 & -2.30 & -5.34 & 14.30$\pm$0.05 &  0.45$\pm$0.10 & 0.40$\pm$0.11 &  8511$\pm$ 281 & 23152$\pm$  765  & 1.25$\pm$0.05   & 0.27$\pm$0.02 \\ 
blg145 & -2.02 & -5.84 & 14.21$\pm$0.01 &  0.48$\pm$0.01 & 0.49$\pm$0.02 &  5951$\pm$ 136 & 16209$\pm$  371  & 1.08$\pm$0.03   & 0.22$\pm$0.01 \\ 
blg152D & -1.81 & -5.31 & 14.26$\pm$0.11 &  0.38$\pm$0.17 & 0.56$\pm$0.20 &  1885$\pm$  88 & 20556$\pm$  958  & 1.08$\pm$0.07   & 0.29$\pm$0.03 \\ 
blg146 & -1.74 & -6.34 & 14.24$\pm$0.01 &  0.51$\pm$0.01 & 0.48$\pm$0.02 &  4171$\pm$ 119 & 11742$\pm$  336  & 0.96$\pm$0.04   & 0.20$\pm$0.01 \\ 
blg152A & -1.66 & -5.57 & 14.21$\pm$0.03 &  0.43$\pm$0.02 & 0.51$\pm$0.05 &  1549$\pm$  70 & 16885$\pm$  765  & 0.96$\pm$0.06   & 0.22$\pm$0.02 \\ 
blg152C & -1.54 & -5.16 & 14.26$\pm$0.03 &  0.45$\pm$0.02 & 0.42$\pm$0.05 &  2100$\pm$  84 & 22881$\pm$  917  & 1.10$\pm$0.06   & 0.23$\pm$0.02 \\ 
blg152B & -1.39 & -5.42 & 14.26$\pm$0.08 &  0.48$\pm$0.10 & 0.45$\pm$0.13 &  1886$\pm$ 125 & 20503$\pm$ 1353  & 1.19$\pm$0.10   & 0.28$\pm$0.03 \\ 
blg153 & -1.32 & -5.86 & 14.24$\pm$0.01 &  0.47$\pm$0.01 & 0.44$\pm$0.02 &  5516$\pm$ 131 & 15125$\pm$  358  & 1.04$\pm$0.03   & 0.21$\pm$0.01 \\ 
blg159D & -1.32 & -4.96 & 14.27$\pm$0.03 &  0.50$\pm$0.02 & 0.44$\pm$0.04 &  2549$\pm$ 100 & 27553$\pm$ 1077  & 1.31$\pm$0.07   & 0.25$\pm$0.02 \\ 
blg159A & -1.18 & -5.22 & 14.37$\pm$0.02 &  0.73$\pm$0.07 & 0.13$\pm$0.03 &  2151$\pm$ 116 & 23282$\pm$ 1259  & 1.28$\pm$0.09   & 0.31$\pm$0.02 \\ 
blg159C & -1.06 & -4.81 & 14.27$\pm$0.02 &  0.45$\pm$0.02 & 0.44$\pm$0.03 &  2468$\pm$  85 & 26713$\pm$  919  & 1.15$\pm$0.06   & 0.22$\pm$0.01 \\ 
blg154 & -1.04 & -6.38 & 14.24$\pm$0.01 &  0.52$\pm$0.01 & 0.43$\pm$0.02 &  4006$\pm$ 120 & 11315$\pm$  338  & 1.04$\pm$0.04   & 0.22$\pm$0.01 \\ 
blg159B & -0.91 & -5.07 & 14.23$\pm$0.02 &  0.44$\pm$0.02 & 0.43$\pm$0.04 &  2143$\pm$  79 & 23148$\pm$  850  & 1.17$\pm$0.06   & 0.22$\pm$0.01 \\ 
blg160 & -0.84 & -5.53 & 14.22$\pm$0.01 &  0.49$\pm$0.01 & 0.46$\pm$0.01 &  6672$\pm$ 149 & 18187$\pm$  405  & 1.13$\pm$0.03   & 0.22$\pm$0.01 \\ 
blg161 & -0.56 & -6.02 & 14.21$\pm$0.01 &  0.51$\pm$0.01 & 0.43$\pm$0.02 &  5569$\pm$ 134 & 15189$\pm$  366  & 1.09$\pm$0.04   & 0.22$\pm$0.01 \\ 
blg168D & -0.49 & -5.24 & 14.25$\pm$0.02 &  0.50$\pm$0.02 & 0.39$\pm$0.03 &  2091$\pm$  81 & 22879$\pm$  886  & 1.25$\pm$0.07   & 0.23$\pm$0.01 \\ 
blg168A & -0.34 & -5.50 & 14.26$\pm$0.02 &  0.46$\pm$0.02 & 0.40$\pm$0.04 &  1576$\pm$  68 & 17199$\pm$  745  & 1.05$\pm$0.06   & 0.21$\pm$0.01 \\ 
blg162 & -0.29 & -6.52 & 14.17$\pm$0.01 &  0.53$\pm$0.01 & 0.39$\pm$0.02 &  3419$\pm$ 109 &  9710$\pm$  309  & 0.92$\pm$0.04   & 0.21$\pm$0.01 \\ 
blg168C & -0.22 & -5.09 & 14.22$\pm$0.05 &  0.45$\pm$0.09 & 0.38$\pm$0.12 &  2128$\pm$  81 & 23289$\pm$  882  & 1.20$\pm$0.06   & 0.25$\pm$0.03 \\ 
blg357 & -0.13 & 5.64 & 14.35$\pm$0.01 &  0.49$\pm$0.01 & 0.40$\pm$0.01 &  5933$\pm$ 143 & 17152$\pm$  413  & 1.08$\pm$0.04   & 0.20$\pm$0.01 \\ 
blg168B & -0.07 & -5.35 & 14.24$\pm$0.02 &  0.44$\pm$0.02 & 0.41$\pm$0.03 &  1777$\pm$  74 & 19425$\pm$  805  & 1.12$\pm$0.06   & 0.20$\pm$0.02 \\ 
blg169 & -0.00 & -5.80 & 14.23$\pm$0.01 &  0.49$\pm$0.01 & 0.39$\pm$0.02 &  6485$\pm$ 145 & 18026$\pm$  404  & 1.21$\pm$0.04   & 0.22$\pm$0.01 \\ 
blg176D & 0.03 & -4.95 & 14.22$\pm$0.02 &  0.45$\pm$0.02 & 0.43$\pm$0.03 &  2384$\pm$  82 & 25796$\pm$  891  & 1.25$\pm$0.06   & 0.22$\pm$0.01 \\ 
blg176A & 0.17 & -5.21 & 14.23$\pm$0.02 &  0.45$\pm$0.02 & 0.43$\pm$0.03 &  2063$\pm$  80 & 22171$\pm$  856  & 1.16$\pm$0.06   & 0.21$\pm$0.01 \\ 
blg170 & 0.27 & -6.31 & 14.19$\pm$0.01 &  0.52$\pm$0.01 & 0.36$\pm$0.02 &  3981$\pm$ 117 & 11522$\pm$  337  & 0.90$\pm$0.03   & 0.20$\pm$0.01 \\ 
blg176C & 0.29 & -4.80 & 14.20$\pm$0.02 &  0.46$\pm$0.01 & 0.45$\pm$0.03 &  2578$\pm$  82 & 27762$\pm$  880  & 1.25$\pm$0.06   & 0.22$\pm$0.01 \\ 
blg176B & 0.44 & -5.07 & 14.23$\pm$0.02 &  0.41$\pm$0.01 & 0.43$\pm$0.04 &  2122$\pm$  73 & 22830$\pm$  789  & 1.14$\pm$0.05   & 0.20$\pm$0.01 \\ 
blg177 & 0.51 & -5.51 & 14.23$\pm$0.01 &  0.47$\pm$0.01 & 0.39$\pm$0.01 &  7176$\pm$ 146 & 19496$\pm$  396  & 1.13$\pm$0.03   & 0.20$\pm$0.01 \\ 
blg355 & 0.62 & 5.57 & 14.31$\pm$0.01 &  0.49$\pm$0.01 & 0.39$\pm$0.02 &  6502$\pm$ 158 & 18471$\pm$  448  & 1.12$\pm$0.04   & 0.22$\pm$0.01 \\ 
blg178 & 0.78 & -6.01 & 14.20$\pm$0.01 &  0.48$\pm$0.01 & 0.34$\pm$0.02 &  5229$\pm$ 128 & 14455$\pm$  354  & 1.03$\pm$0.03   & 0.20$\pm$0.01 \\ 
blg186D & 0.81 & -5.14 & 14.24$\pm$0.01 &  0.46$\pm$0.02 & 0.32$\pm$0.03 &  2143$\pm$  79 & 23201$\pm$  855  & 1.13$\pm$0.06   & 0.20$\pm$0.01 \\ 
blg186A & 0.95 & -5.40 & 14.22$\pm$0.01 &  0.48$\pm$0.02 & 0.35$\pm$0.02 &  1886$\pm$  71 & 20401$\pm$  763  & 1.11$\pm$0.06   & 0.18$\pm$0.01 \\ 
blg186C & 1.08 & -4.99 & 14.24$\pm$0.01 &  0.45$\pm$0.02 & 0.36$\pm$0.03 &  2416$\pm$  82 & 26159$\pm$  887  & 1.23$\pm$0.06   & 0.21$\pm$0.01 \\ 
blg186B & 1.22 & -5.25 & 14.25$\pm$0.01 &  0.47$\pm$0.02 & 0.30$\pm$0.03 &  2253$\pm$  77 & 24284$\pm$  832  & 1.30$\pm$0.06   & 0.21$\pm$0.01 \\ 
blg187 & 1.29 & -5.69 & 14.21$\pm$0.01 &  0.46$\pm$0.01 & 0.34$\pm$0.02 &  5933$\pm$ 134 & 16256$\pm$  368  & 1.04$\pm$0.03   & 0.20$\pm$0.01 \\ 
blg192 & 1.65 & -5.12 & 14.23$\pm$0.01 &  0.43$\pm$0.01 & 0.32$\pm$0.02 &  8579$\pm$ 160 & 23301$\pm$  434  & 1.15$\pm$0.03   & 0.20$\pm$0.01 \\ 
blg193 & 1.92 & -5.62 & 14.21$\pm$0.01 &  0.45$\pm$0.01 & 0.32$\pm$0.02 &  6390$\pm$ 138 & 17606$\pm$  382  & 1.09$\pm$0.03   & 0.20$\pm$0.01 \\ 
blg199 & 2.44 & -5.33 & 14.21$\pm$0.01 &  0.44$\pm$0.02 & 0.26$\pm$0.02 &  8183$\pm$ 156 & 22176$\pm$  424  & 1.20$\pm$0.03   & 0.21$\pm$0.01 \\ 
blg204 & 2.71 & -5.84 & 14.15$\pm$0.01 &  0.46$\pm$0.01 & 0.27$\pm$0.02 &  6216$\pm$ 140 & 17163$\pm$  386  & 1.16$\pm$0.03   & 0.21$\pm$0.01 \\ 
blg211 & 3.03 & -5.18 & 14.20$\pm$0.01 &  0.44$\pm$0.02 & 0.25$\pm$0.02 &  8144$\pm$ 158 & 22394$\pm$  435  & 1.21$\pm$0.03   & 0.22$\pm$0.01 \\ 
blg212 & 3.30 & -5.68 & 14.18$\pm$0.02 &  0.45$\pm$0.03 & 0.21$\pm$0.04 &  6966$\pm$ 147 & 18908$\pm$  399  & 1.22$\pm$0.03   & 0.24$\pm$0.01 \\ 
blg213 & 3.57 & -6.19 & 14.14$\pm$0.01 &  0.46$\pm$0.02 & 0.24$\pm$0.03 &  5216$\pm$ 130 & 14633$\pm$  364  & 1.12$\pm$0.04   & 0.22$\pm$0.01 \\ 
blg221 & 3.84 & -5.42 & 14.14$\pm$0.01 &  0.39$\pm$0.02 & 0.26$\pm$0.04 &  7375$\pm$ 147 & 20022$\pm$  399  & 1.17$\pm$0.03   & 0.21$\pm$0.01 \\ 
\hline\hline
\end{tabular}
\label{table:DoubleRCparameters}
\end{table*}

\begin{table*}
\normalsize
\caption{\large Parameters of split red clumps toward 33 additional fields not used for calibration. Definitions of the parameters and can be found in the caption to Table \ref{table:DoubleRCparameters}. The larger error values and decreased trends demonstrate that the split red clump is sharply falling off toward these fields, which are at larger longitudinal separation from the Galactic minor axis than the calibration fields.} 
\centering 
\begin{tabular}{llllllllll}
\hline\hline\hline 
Field & $l$ & $b$ & $I_{RC,1}$ & ${\Delta}I$ & $N_{\rm{Faint}}/N_{\rm{Total}}$ & $N_{RC}$ & $\Sigma_{RC}$   & $EW_{RC}$ & $\sigma_{RC,1}$ \\
 \hline\hline
blg327 & -9.17 & -5.29 & 14.54$\pm$0.19 &  0.48$\pm$0.17 & 0.19$\pm$0.25 &  1755$\pm$ 110 &  5284$\pm$  331  & 0.53$\pm$0.04   & 0.24$\pm$0.04 \\ 
blg328 & -8.87 & -5.79 & 14.24$\pm$0.19 &  0.46$\pm$0.17 & 0.80$\pm$0.24 &  1656$\pm$ 116 &  4859$\pm$  341  & 0.52$\pm$0.05   & 0.32$\pm$0.04 \\ 
blg329 & -8.59 & -6.27 & 14.46$\pm$0.02 &  0.36$\pm$0.02 & 0.43$\pm$0.04 &   862$\pm$  57 &  3769$\pm$  249  & 0.48$\pm$0.04   & 0.15$\pm$0.02 \\ 
blg322 & -8.57 & -5.28 & 14.55$\pm$0.04 &  0.58$\pm$0.12 & 0.13$\pm$0.09 &  1900$\pm$ 119 &  7564$\pm$  472  & 0.70$\pm$0.05   & 0.27$\pm$0.02 \\ 
blg323 & -8.41 & -5.72 & 14.30$\pm$0.09 &  0.43$\pm$0.17 & 0.56$\pm$0.17 &   457$\pm$  42 &  6226$\pm$  571  & 0.65$\pm$0.08   & 0.20$\pm$0.04 \\ 
blg324 & -7.94 & -6.24 & 14.59$\pm$0.07 &  0.00$\pm$0.17 & 0.00$\pm$0.14 &  2065$\pm$ 124 &  6054$\pm$  364  & 0.68$\pm$0.05   & 0.31$\pm$0.04 \\ 
blg319 & -7.86 & -5.21 & 14.57$\pm$0.04 &  0.67$\pm$0.12 & 0.09$\pm$0.10 &  3395$\pm$ 193 &  9430$\pm$  535  & 0.78$\pm$0.05   & 0.29$\pm$0.02 \\ 
blg320 & -7.57 & -5.71 & 14.54$\pm$0.06 &  0.79$\pm$0.17 & 0.05$\pm$0.15 &  3280$\pm$ 181 &  9014$\pm$  497  & 0.85$\pm$0.05   & 0.30$\pm$0.03 \\ 
blg321 & -7.28 & -6.19 & 14.55$\pm$0.06 &  0.80$\pm$0.17 & 0.05$\pm$0.13 &  2589$\pm$ 161 &  7867$\pm$  489  & 0.81$\pm$0.06   & 0.30$\pm$0.03 \\ 
blg318 & -7.02 & -5.46 & 14.50$\pm$0.05 &  0.41$\pm$0.17 & 0.22$\pm$0.21 &  3741$\pm$ 155 & 10269$\pm$  425  & 0.82$\pm$0.04   & 0.25$\pm$0.04 \\ 
blg315 & -6.36 & -5.42 & 14.37$\pm$0.03 &  0.38$\pm$0.02 & 0.52$\pm$0.04 &  4583$\pm$ 128 & 12869$\pm$  358  & 0.93$\pm$0.03   & 0.21$\pm$0.01 \\ 
blg110 & -5.85 & -5.12 & 14.40$\pm$0.03 &  0.37$\pm$0.02 & 0.48$\pm$0.07 &  6958$\pm$ 160 & 18892$\pm$  433  & 1.06$\pm$0.03   & 0.22$\pm$0.02 \\ 
blg111 & -5.56 & -5.61 & 14.33$\pm$0.02 &  0.41$\pm$0.02 & 0.57$\pm$0.03 &  4020$\pm$ 109 & 15475$\pm$  421  & 1.07$\pm$0.04   & 0.22$\pm$0.01 \\ 
blg115 & -5.42 & -5.22 & 13.90$\pm$0.12 &  0.79$\pm$0.10 & 0.74$\pm$0.06 &   157$\pm$  23 & 25301$\pm$ 3707  & 1.37$\pm$0.30   & 0.30$\pm$0.05 \\ 
blg336 & 4.50 & 5.06 & 14.24$\pm$0.11 &  0.71$\pm$0.11 & 0.08$\pm$0.08 &  4356$\pm$  61 & 23591$\pm$  331  & 1.30$\pm$0.02   & 0.28$\pm$0.01 \\ 
blg230 & 4.54 & -5.48 & 14.12$\pm$0.02 &  0.41$\pm$0.02 & 0.26$\pm$0.04 &  4646$\pm$ 115 & 17021$\pm$  420  & 1.10$\pm$0.04   & 0.21$\pm$0.01 \\ 
blg231 & 4.80 & -5.98 & 14.13$\pm$0.02 &  0.41$\pm$0.03 & 0.25$\pm$0.05 &  5383$\pm$ 131 & 15267$\pm$  372  & 1.12$\pm$0.04   & 0.22$\pm$0.01 \\ 
blg239 & 5.13 & -5.34 & 14.13$\pm$0.02 &  0.39$\pm$0.03 & 0.25$\pm$0.06 &  6430$\pm$ 140 & 17661$\pm$  384  & 1.13$\pm$0.03   & 0.23$\pm$0.01 \\ 
blg240 & 5.40 & -5.84 & 14.17$\pm$0.13 &  0.10$\pm$0.13 & 0.50$\pm$0.22 &  3582$\pm$ 106 & 14014$\pm$  415  & 1.04$\pm$0.04   & 0.32$\pm$0.03 \\ 
blg246 & 5.59 & -5.21 & 14.10$\pm$0.03 &  0.44$\pm$0.13 & 0.25$\pm$0.20 &  1835$\pm$  72 & 20361$\pm$  801  & 1.33$\pm$0.07   & 0.24$\pm$0.04 \\ 
blg247 & 6.01 & -5.72 & 14.11$\pm$0.02 &  0.43$\pm$0.02 & 0.30$\pm$0.04 &  5194$\pm$ 125 & 14146$\pm$  340  & 1.12$\pm$0.04   & 0.22$\pm$0.01 \\ 
blg248 & 6.28 & -6.21 & 13.74$\pm$0.14 &  0.47$\pm$0.14 & 0.91$\pm$0.13 &  3601$\pm$ 114 & 10667$\pm$  337  & 1.02$\pm$0.04   & 0.33$\pm$0.01 \\ 
blg255 & 6.52 & -5.42 & 14.00$\pm$0.08 &  0.41$\pm$0.09 & 0.48$\pm$0.16 &  5792$\pm$ 144 & 15804$\pm$  393  & 1.19$\pm$0.04   & 0.29$\pm$0.02 \\ 
blg256 & 6.79 & -5.93 & 14.08$\pm$0.08 &  0.43$\pm$0.17 & 0.25$\pm$0.21 &  4298$\pm$ 157 & 11930$\pm$  436  & 1.08$\pm$0.05   & 0.26$\pm$0.04 \\ 
blg262 & 7.04 & -5.11 & 14.00$\pm$0.02 &  0.46$\pm$0.01 & 0.42$\pm$0.04 &  5388$\pm$ 134 & 14766$\pm$  367  & 1.14$\pm$0.04   & 0.25$\pm$0.02 \\ 
blg263 & 7.30 & -5.63 & 14.18$\pm$0.09 &  0.11$\pm$0.17 & 0.27$\pm$0.20 &  4863$\pm$ 142 & 13210$\pm$  386  & 1.07$\pm$0.04   & 0.35$\pm$0.03 \\ 
blg264 & 7.57 & -6.13 & 14.17$\pm$0.08 &  0.79$\pm$0.14 & 0.03$\pm$0.20 &  3666$\pm$ 111 & 10427$\pm$  314  & 1.10$\pm$0.04   & 0.30$\pm$0.04 \\ 
blg268 & 7.66 & -5.03 & 13.99$\pm$0.03 &  0.47$\pm$0.02 & 0.44$\pm$0.05 &  5269$\pm$ 141 & 14474$\pm$  387  & 1.05$\pm$0.04   & 0.26$\pm$0.02 \\ 
blg269 & 7.93 & -5.55 & 13.99$\pm$0.06 &  0.40$\pm$0.17 & 0.46$\pm$0.14 &  3935$\pm$ 118 & 10825$\pm$  325  & 0.90$\pm$0.03   & 0.25$\pm$0.02 \\ 
blg270 & 8.19 & -6.06 & 13.93$\pm$0.07 &  0.41$\pm$0.12 & 0.48$\pm$0.15 &  3240$\pm$ 109 &  8957$\pm$  301  & 0.94$\pm$0.04   & 0.26$\pm$0.03 \\ 
blg275 & 8.34 & -5.05 & 13.98$\pm$0.05 &  0.47$\pm$0.03 & 0.38$\pm$0.09 &  4161$\pm$ 130 & 12311$\pm$  384  & 1.00$\pm$0.04   & 0.28$\pm$0.02 \\ 
blg271 & 8.47 & -6.55 & 14.12$\pm$0.14 &  0.16$\pm$0.17 & 0.14$\pm$0.22 &  1679$\pm$  77 &  6995$\pm$  320  & 0.86$\pm$0.05   & 0.30$\pm$0.04 \\ 
blg276 & 8.55 & -5.59 & 14.03$\pm$0.10 &  0.47$\pm$0.17 & 0.22$\pm$0.21 &  1747$\pm$ 107 &  9589$\pm$  586  & 0.90$\pm$0.06   & 0.28$\pm$0.04 \\ 
\hline\hline
\end{tabular}
\label{table:DoubleRCparametersFail} 
\end{table*}

\twocolumn 

\begin{figure*}
\begin{center}
\includegraphics[totalheight=0.36\textheight]{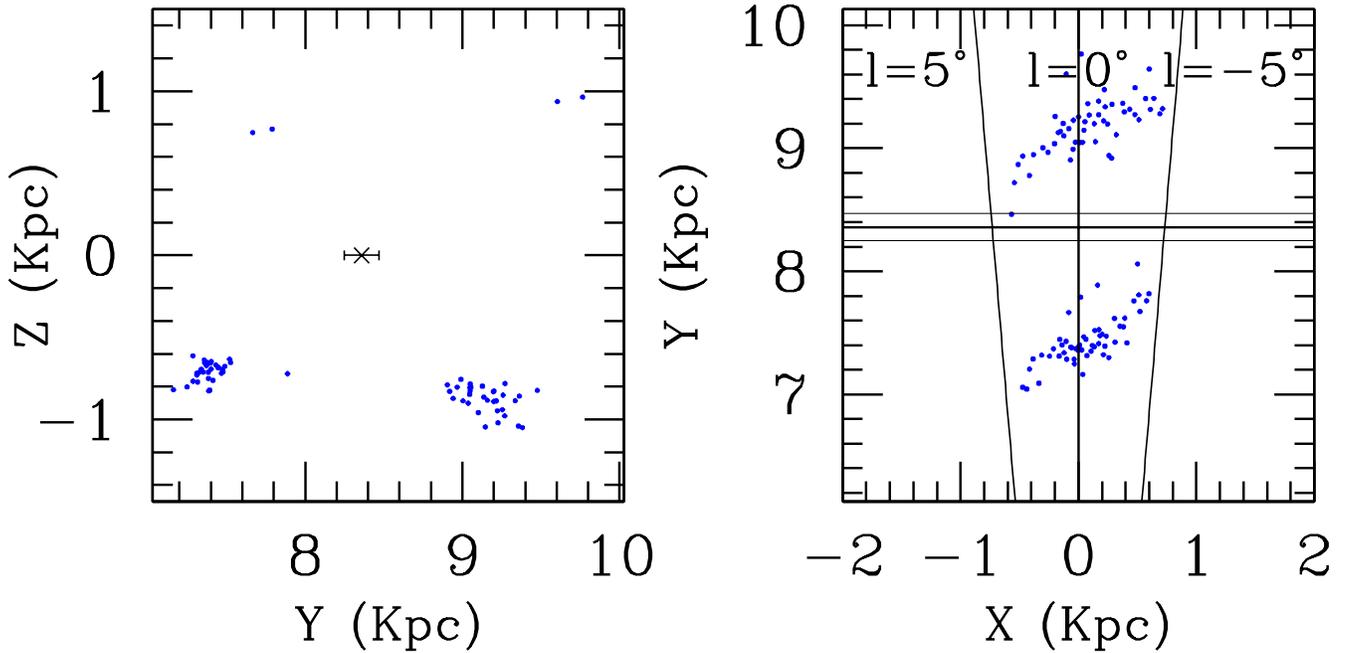}
\end{center}
\caption{\large The mean positions of the observed red clump centroids in Galactic coordinate coordinates, where the Sun is located at $(X,Y,Z)=(0,0,0)$ and positive longitudes denote negative X-directions. The positions are derived by converting the mean brightness of the red clump into a distance by assuming an absolute magnitude of $M_{I,RC} = -0.12$. We show the estimated position of the Galactic centre with 1-$\sigma$ error bars, $(8.36 \pm 0.11)$ Kpc, from \citet{2014arXiv1403.5266C}, on both panels.  : Projection of the red clump centroids onto the XZ plane, for measurements satisfying $-1^{\circ} \leq l \leq 1^{\circ}$. RIGHT: Projection of the RC centroids onto the XY plane. }
\label{Fig:BrightnessPlot4Paper}
\end{figure*}

\subsection{The Brightness of the Two Red Clumps as a Function of Position}
\label{sec:BrightnessResults}

The apparent magnitude of the brighter RC is a decreasing function of increasing longitude and increasing separation from the plane:
\begin{equation}
\begin{split}
I_{RC,1} = (14.263 \pm 0.004)+ (-0.022 \pm 0.001){\times}(l) \\ 
+(-0.042 \pm 0.004){\times}(|b|-5), \\
\delta = 0.0034, \\
\chi^2 = 78.22, \\
\rm{DoF} = 43,
\end{split}
\end{equation}
where the fit (and all subsequent linear fits in this section) is computed using the standard formalism for weighted multilinear least-squares fit, with the weights given by the inverse square of the errors on the measured values of the dependent variable (in this case $I_{RC,1} $), and 3.0$\sigma$ outliers recursively removed. We also state the resultant scatter $\delta$ and $\chi^2$, as well as the degrees of freedom (DoF) of the fit, which is equal to the difference between the number of non-outlier data points and the number of free parameters to the fit.

The brightness separation between the two RCs is a slowly and quadratically decreasing function of separation from the minor axis and a rapidly increasing function of separation from the plane:
\begin{equation}
\begin{split}
{\Delta}I = (0.449 \pm 0.004) + (-0.0049 \pm 0.0003){\times}(l)^2  \\ 
+ (0.053 \pm 0.004){\times}(|b|-5), \\
\delta = 0.0025, \\
\chi^2 = 48.22, \\
\rm{DoF} = 48.
\end{split}
\label{EQ:DeltaI}
\end{equation}
The latitudinal dependence clearly illustrates the fact that at larger separations from the plane, the arms of the X-shape move outward. If we extrapolate the fit to the origin, we obtain a brightness separation toward $(l,b)=(0^{\circ},0^{\circ})$ of ${\Delta}I=0.182\pm0.024$ -- not consistent with zero. Our findings are therefore marginally suggestive of an open-X rather than an closed-X morphology for the Galactic bulge, in other words the extrapolations of the arms of the X-shape, were they to be viewed side on, would not intersect at the Galactic centre, see \citet{2006MNRAS.370..753B} and \citet{2002MNRAS.337..578P} for further discussion.  

We show the projected positions of the RC centroids (assuming $M_{I,RC}=-0.12$, \citealt{2013ApJ...769...88N}) in Figure \ref{Fig:BrightnessPlot4Paper}.

\begin{figure*}
\begin{center}
\includegraphics[totalheight=0.35\textheight]{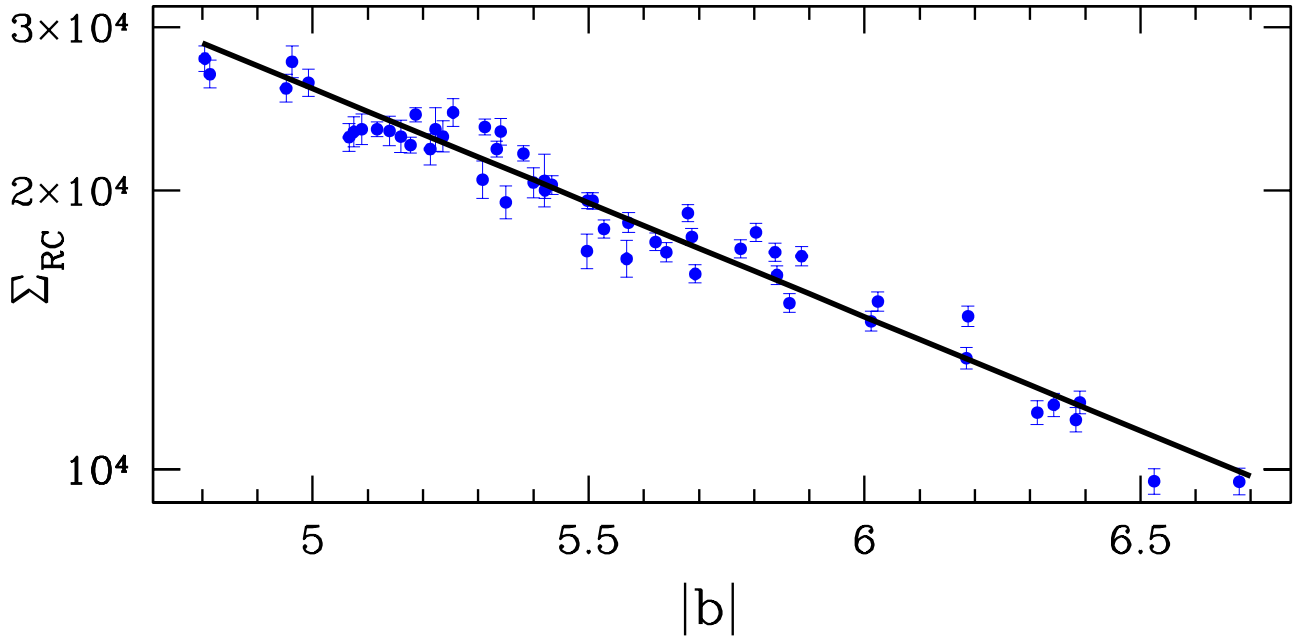}
\end{center}
\caption{\large The surface density of red clump stars (in degrees$^{-2}$), $\Sigma_{RC}$, as a function of absolute latitude $|b|$. The data from individual fields are shown in blue, and the best-fit relation from Equation \ref{EQ:SurfaceDensity} is shown in black.  } 
\label{Fig:DensityPlot4Paper}
\end{figure*}

\subsection{The Combined Surface Density of Red Clump Stars on the Sky}
\label{sec:DensityResults}

The combined surface density of bright and faint RC stars on the sky is a steeply declining function of separation from the plane:
\begin{equation}
\begin{split}
\log_{10} ({\Sigma_{RC}}) = (4.411 \pm 0.003) + (-0.246 \pm 0.004){\times}(|b|-5), \\
\delta = 0.0035, \\
\chi^2 = 97.28, \\
\rm{DoF} = 36.
\end{split}
\label{EQ:SurfaceDensity}
\end{equation}
This can also be discerned in Figure  \ref{Fig:DensityPlot4Paper}, where we show the scatter of surface density as a function of absolute latitude. The surface density - latitude relation is strikingly consistent with an exponential density profile for the Galactic bulge.

\subsection{The Fraction of Red Clump Stars in the Faint Red Clump}
\label{sec:DensityRatio}

\begin{figure*}
\begin{center}
\includegraphics[totalheight=0.30\textheight]{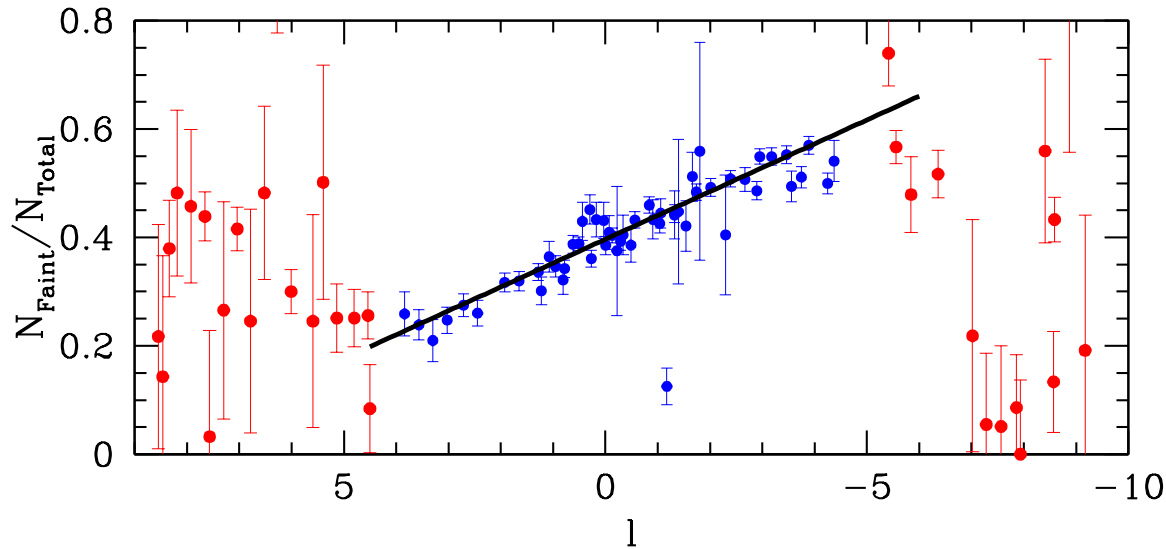}
\end{center}
\caption{\large The fraction of red clump stars in the faint red clump, $N_{\rm{Faint}}/N_{\rm{Total}}$, is plotted as a function of Galactic longitude. The measurements for the calibration fields are shown in blue, and their best-fit relation is shown in black. The measurements for 33 additional fields are shown in red. } 
\label{Fig:FractionPlot4Paper}
\end{figure*}

The best-fit relation for the fraction of stars in the fainter RC is:
\begin{equation}
\begin{split}
N_{\rm{Faint}}/N_{\rm{Total}} = (0.402 \pm 0.055)  + (-0.044 \pm 0.002){\times}(l) \\
  + (-0.010 \pm 0.007){\times}(|b|-5), \\
\delta = 0.0047, \\
\chi^2 = 56.83, \\
\rm{DoF} = 49.
\end{split}
\end{equation}
The scatter of  of $N_{\rm{Faint}}/N_{\rm{Total}}$ vs $l$ is shown in Figure \ref{Fig:FractionPlot4Paper}. 

The reason for this correlation is that the Galactic bar's major axis is at an angle relative to the Sun-Galactic Centre line of sight \citep{1997ApJ...477..163S,2005MNRAS.358.1309B,2013MNRAS.434..595C,2013MNRAS.435.1874W}. If the arms of the X-shape protrude outward from the Galactic bar, it logically follows  that the brighter RC should be relatively less numerous toward more negative longitudes, as the brighter RC would protrude from the nearer side of the bar. That the measurements become more erratic toward larger angular separations from the Galactic minor axis (red points, Figure \ref{Fig:FractionPlot4Paper}) is possibly an indication of the ``end" of the X-shape feature.  N-body simulations consistently predict that the length of the peanut should be shorter than the total length of the bar closer to the plane \citep{2005MNRAS.358.1477A}, and this was specifically argued for our Galaxy by \citet{2011ApJ...734L..20M} and \citet{2011MNRAS.418.1176R}. 

\begin{figure}
\begin{center}
\includegraphics[totalheight=0.35\textheight]{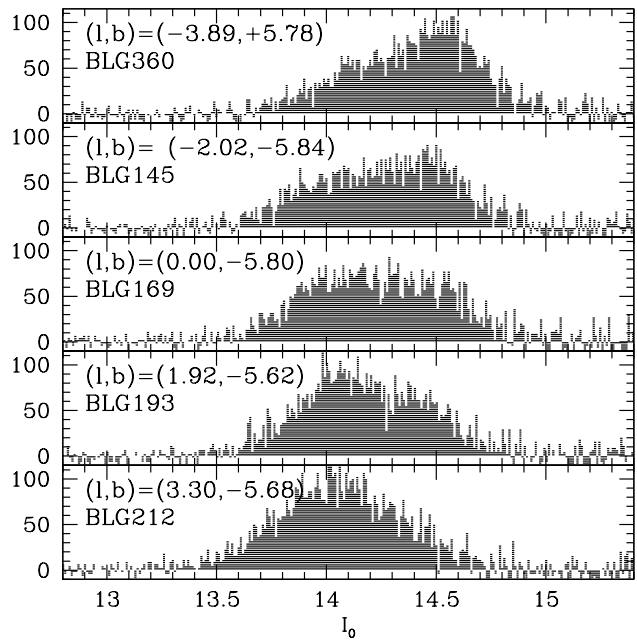}
\end{center}
\caption{\large The fraction of RC stars in the faint-RC is a decreasing function of longitude. Shown are the dereddened magnitude histograms for five OGLE-III fields with the continuum fits to the RG+RGBB+AGBB removed from the observed number counts. Each histogram covers the same viewing area and is shown on the same scale. The best-fit ratio $N_{\rm{Faint}}/N_{\rm{Total}}$ progressively decreases from $(57\pm2)$\% toward $(l,b)=(-3.89^{\circ},+5.78^{\circ})$ to $(21\pm4)$\% toward $(l,b)=(3.30^{\circ},-5.68^{\circ})$. With the continuum RG+RGBB+AGB components to the luminosity function statistically subtracted, it is clear that the two peaks are slightly skew-symmetric.} 
\label{Fig:LongitudeHistogram}
\end{figure}

 In Figure \ref{Fig:LongitudeHistogram}, we show the observed magnitude distribution of RC stars for five fields with varying longitude but nearly constant latitude. The fainter RC is virtually invisible toward $(l,b)=(3.30^{\circ},-5.68^{\circ})$ (field BLG 212) and becomes progressively clearer as longitude is decreased.  
 
\subsection{The Ratio of Red Clump Stars to Red Giant Stars}
\label{sec:EWRC}

In Section \ref{sec:RCmeasurements}, we defined $EW_{RC}$ to be the ratio of the total number of RC stars to number density of RG stars per magnitude at the luminosity of the RC. We find that $EW_{RC}$ is a slowly declining function of decreasing longitude and a steeply declining function of increasing absolute latitude (previously noted by \citealt{2012A&A...546A..57U}, see their Figure 11 and related discussion):
\begin{equation}
\begin{split}
EW_{RC} = (1.213 \pm 0.010)+ (0.007 \pm 0.002){\times}(l) \\ 
+ (-0.164 \pm 0.013){\times}(|b|-5), \\
\delta = 0.0083, \\
\chi^2 = 78.01, \\
\rm{DoF} = 50.
\end{split}
\end{equation}
The mean value is significantly below the intrinsic value of $EW_{RC} \approx 2.10$ \citep{2014MNRAS.442.2075N}. At least a part of this is likely due to disk contamination: foreground RGs will have the same apparent magnitude as the RC if they are intrinsically fainter, which is where the luminosity function of RGs is highest. In contrast, the median value predicted by the convolution of the theoretical luminosity function with the N-body model is $EW_{RC} \approx 1.75$ for the models of \citet{2012ApJ...756...22N} and $EW_{RC}\approx 1.52$ for the model of \citet{2010ApJ...720L..72S}. 

The solution to this discrepancy that we consider the likeliest is that the rate of disk contamination is higher than the rates of $\sim$6\% and $\sim$11\% predicted by the respective N-models of \citet{2012ApJ...756...22N} and \citet{2010ApJ...720L..72S}  -- there is no reason to assume that the bulge-to-disk ratio in the model and their distributions along the line of sight would be the same as that for the Galaxy. A reduction from $EW_{RC}=2.10$ to $EW_{RC}=1.21$ (the value at $|b|=5.0^{\circ}$) will occur if 19\% of the stars near the luminosity of the RC are not bulge stars, and to $EW_{RC}=0.97$ (the value at $|b|=6.5^{\circ}$) will occur if 27\% of the stars near the apparent magnitude of the RC are not bulge stars. These are compatible with the estimate of $\sim 25\%$ derived by \citet{2013MNRAS.430..836N}. Other possibilities include stellar models not correctly predicting the ratio of RC to RG stars, or that a significant fraction of bulge stars never go through the RC phase, see  \citet{2014arXiv1406.6451C}, who found that $\sim$30\% of white dwarfs being systematically cooler than canonical tracks. 

\section{Data and Models: A Comparison of Three Observables}
We compare the model predictions to the data for three observables, and for orientation angles $\alpha_{\rm{Bar}}=0,15,30,\,\rm{and\,}45^{\circ}$. The central coordinates of the model prediction span longitudinal range $l =-10.5,-9.5, ... +9.5$ and the latitudinal range $b=-5.5,-6.5$, with each point corresponding to a square with radius of length of 1 degree.  The root-mean-square of the difference between the spline-interpolated model predictions and the observed parameters in the OGLE-III photometry are listed in Table \ref{table:ModelDataComparison}.

\begin{figure*}
\begin{center}
\includegraphics[totalheight=0.7\textheight]{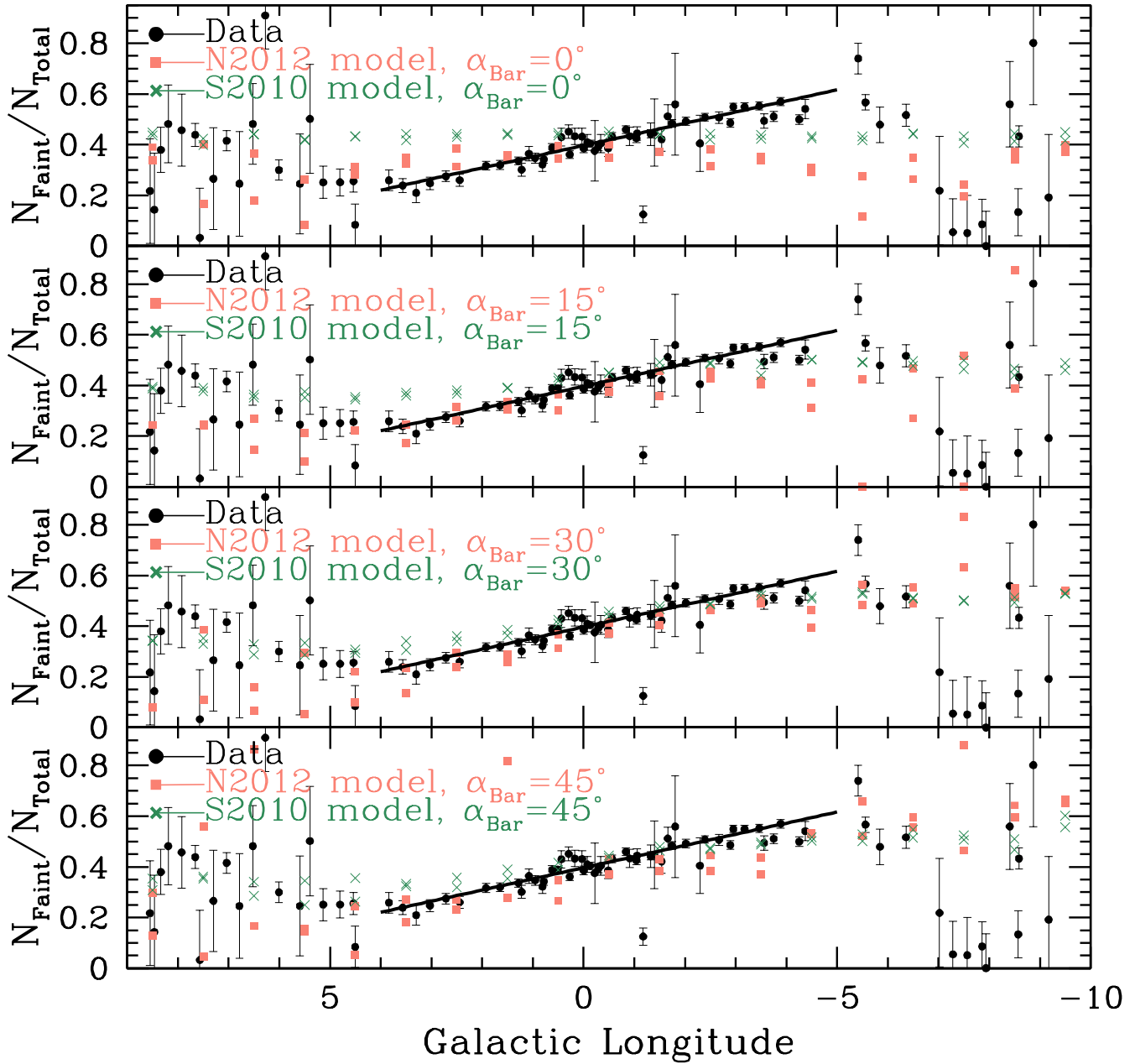}
\end{center}
\caption{\large The fraction of red clump stars in the faint red clump as a function of longitude. Data points and best-fit relation are shown as black circles, predictions from the model are shown as salmon squares \citep{2012ApJ...756...22N} and seagreen X's \citep{2010ApJ...720L..72S}, for four different viewing angles to the Galactic bar, $\alpha_{\rm{Bar}}=0,15,30,45^{\circ}$. The two predictions per N-body model per bin in longitude correspond to the $b=-5.5^{\circ}$ and $b=-6.5^{\circ}$ sightlines.}
\label{Fig:FractionPlot4PaperNbodyAnglesFraction}
\end{figure*}

The fraction of stars in the faint RC as a function of longitude (Figure \ref{Fig:FractionPlot4PaperNbodyAnglesFraction}) is predicted to show an increasing fraction towards decreasing longitudes, as expected, with the obvious exception of the case where the bar's viewing angle is $\alpha_{\rm{Bar}}=0^{\circ}$. The cases $\alpha_{\rm{Bar}}=0,15^{\circ}$ are the least consistent with the observations for both models.  The model of \citet{2012ApJ...756...22N} is most consistent with the case $\alpha_{\rm{Bar}}=30^{\circ}$, whereas that of \citet{2010ApJ...720L..72S} is comparably consistent with the cases $\alpha_{\rm{Bar}}=30,45^{\circ}$ and nearly as consistent with the case $\alpha_{\rm{Bar}}=15^{\circ}$. 

\begin{figure*}
\begin{center}
\includegraphics[totalheight=0.7\textheight]{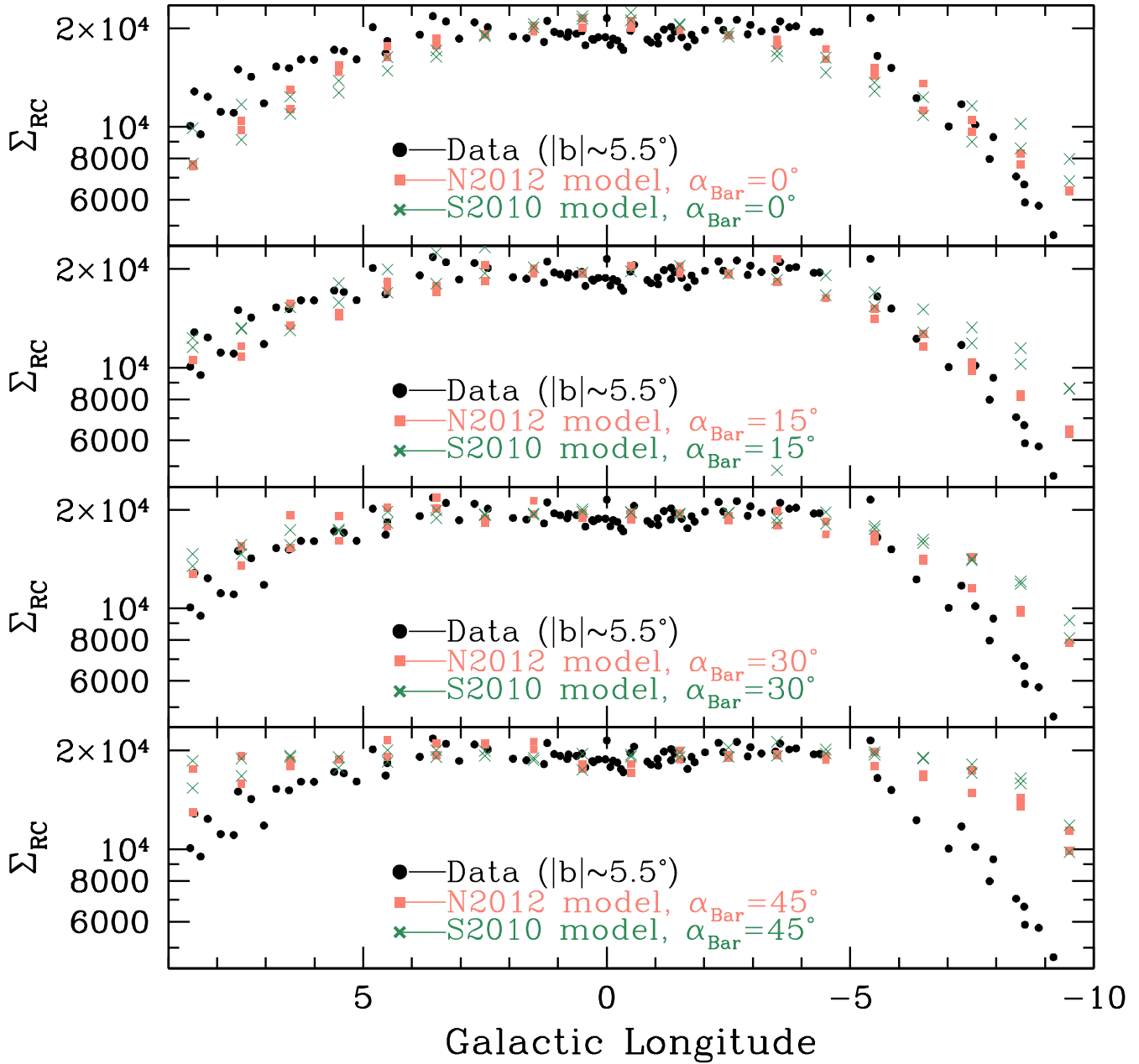}
\end{center}
\caption{\large The surface density of red clump stars in degrees$^{-2}$ as a function of longitude. Data points and best-fit relation are shown as black circles, which  have been normalised to the value at $|b|=5.5^{\circ}$ using Equation \ref{EQ:SurfaceDensity}. Predictions from the model are shown as salmon squares \citep{2012ApJ...756...22N} and seagreen X's \citep{2010ApJ...720L..72S}, for four different viewing angles to the Galactic bar, $\alpha_{\rm{Bar}}=0,15,30,45^{\circ}$. The two predictions per N-body model per bin in longitude correspond to the $b=-5.5^{\circ}$ and $b=-6.5^{\circ}$ sightlines.} 
\label{Fig:FractionPlot4PaperNbodyAnglesDensity}
\end{figure*}

For the surface density of RC stars as a function of longitude (Figure \ref{Fig:FractionPlot4PaperNbodyAnglesDensity}), the data show an interesting feature: a virtually flat surface density in the range $(-5.0^{\circ} \lesssim l \lesssim 4.0^{\circ})$, with a steep log-linear dropoff outside of that range. We have found in our investigation that the combined surface density of RC stars is virtually insensitive to measurement systematics and degeneracies -- the total RC+RGBB number density is a well-behaved observable with minimal uncertainties, and the only remaining degeneracies are how the relevant stars are partitioned into the different groups (bright vs faint RC, etc), not their total normalisation. Indeed, the top-left panel of Figure 21 of \citet{2013ApJ...769...88N}, shows nearly the same profile for $\Sigma_{RC}$ vs $l$ in the case $|b|=5.5^{\circ}$, even though those estimates were obtained using a single RC fit.  We suggest that the fact the number density of stars falls off sharply beyond that longitudinal range may be an indicator of the length of the X-shape feature, at a height corresponding to $|b| \sim5.5^{\circ}$. In other words, if we assume a Galacticentric distance $R_{GC} = 8.36$ Kpc and a viewing angle toward the bar of $\alpha_{\rm{Bar}}=30^{\circ}$, the drop off in density toward $(l,b)=(4^{\circ},|5.5|^{\circ})$ corresponds to a height of 0.72 Kpc from the Galactic plane at a distance of 1.04 Kpc from the Galactic centre.  Conversely, the drop off in density toward $(l,b)=(-5^{\circ},|5.5|^{\circ})$ corresponds to a height of 0.95 Kpc from the Galactic plane at a distance of 1.72 Kpc form the Galactic centre. The first value is consistent with the location of the peak density at that separation from the plane inferred from Figure 19 of \citet{2013MNRAS.435.1874W}, whereas the second value is $\sim$40\% further out, where the density is projected to have dropped closer to its value along the minor axis. 

A peculiarity of this observable is that the predictions of both N-body models relative to the data show the the least integrated scatter for  $\alpha_{\rm{Bar}}=0^{\circ}$. This orientation angle is already ruled out by various other arguments \citep{2013MNRAS.434..595C,2013MNRAS.435.1874W}, and so this trend is unexpected. Inspection of Figure \ref{Fig:FractionPlot4PaperNbodyAnglesDensity} shows that the discrepancy is largely due to predictions in the range $(-10.5^{\circ} \lesssim l \lesssim -5.0^{\circ})$. Figure \ref{Fig:FractionPlot4PaperNbodyAnglesDensity} makes clear that larger values of $\alpha_{\rm{Bar}}$ would be favoured if the comparison were restricted to sightlines closer to the minor axis, which can also be found listed in Table \ref{table:ModelDataComparison}, where we have included the scatter when restricting to sightlines that are closer to the minor axis. 

\begin{figure*}
\begin{center}
\includegraphics[totalheight=0.7\textheight]{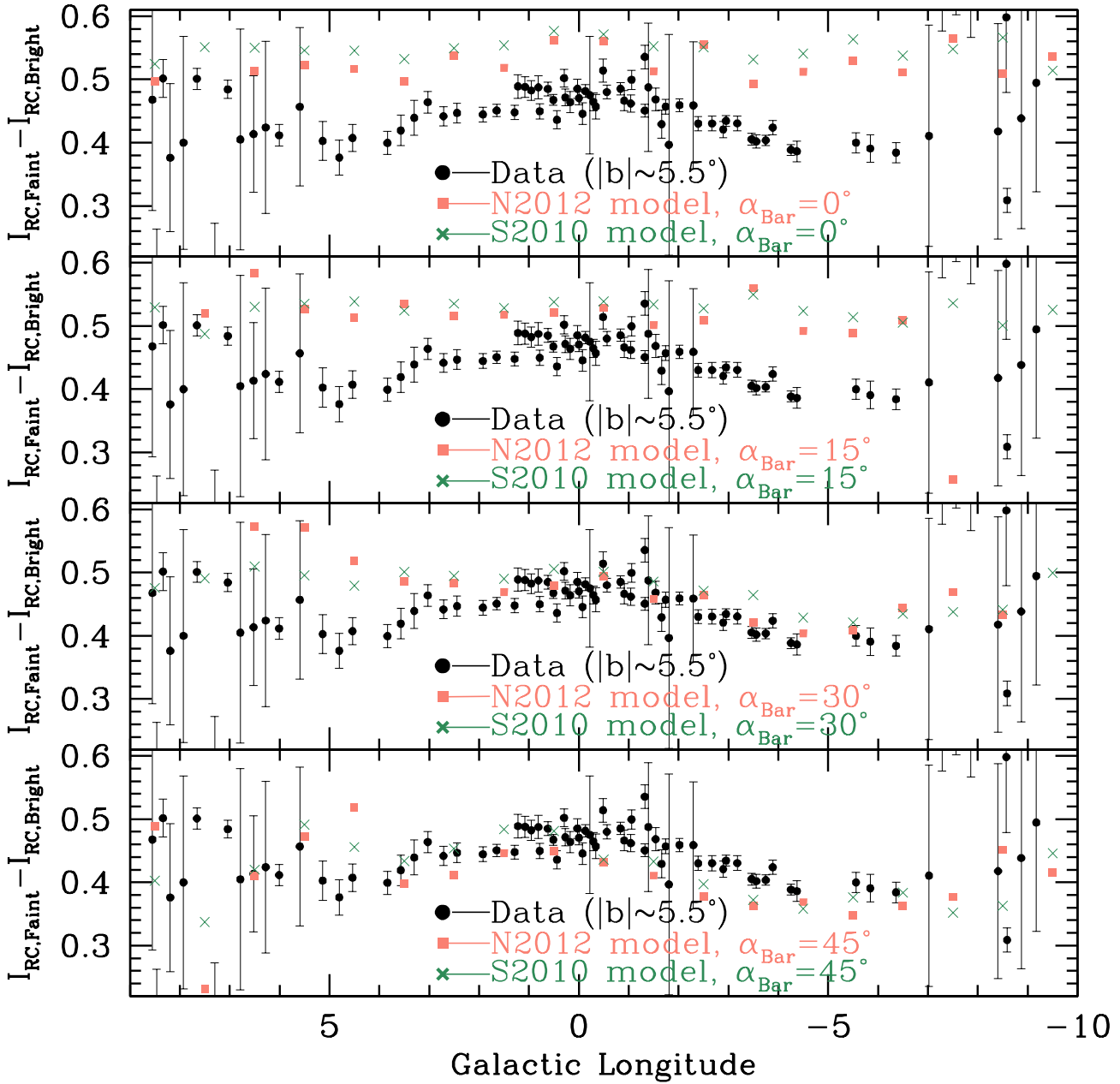}
\end{center}
\caption{\large The brightness difference between the bright and faint red clump as a function of longitude. Data points and best-fit relation are shown as black circles,  which  have been normalised to the value at $|b|=5.5^{\circ}$ using Equation \ref{EQ:DeltaI}. Predictions from the model are shown as salmon squares \citep{2012ApJ...756...22N} and seagreen X's \citep{2010ApJ...720L..72S}, for four different viewing angles to the Galactic bar, $\alpha_{\rm{Bar}}=0,15,30,45^{\circ}$.} 
\label{Fig:DeltaI}
\end{figure*}

The third observable investigated, that of $I_{RC,\rm{Faint}}- I_{RC,\rm{Bright}}$ as a function of longitude, has its observational and theoretical scatter vs longitude shown in Figure \ref{Fig:DeltaI}.  As with the observable $N_{\rm{Faint}}/N_{\rm{Total}}$, the best match for the model of \citet{2012ApJ...756...22N} is for the case  $\alpha_{\rm{Bar}}=30^{\circ}$, whereas the model of  \citet{2010ApJ...720L..72S} yields a comparably good match for the cases $\alpha_{\rm{Bar}}=30,45^{\circ}$.

\begin{table}
\caption{\large Comparison of split red clump parameters in the OGLE-III photometry and in N-body model predictions. For each entry in the table we list the root-mean-square of the spline-interpolated model prediction and the observed value toward the sightline of the observation. The comparison for $N_{\rm{Faint}}/N_{\rm{Total}}$ and ${\Delta}I$ is for the 53 calibration fields that are not BLG 159A (an outlier in all aspects, see Table \ref{table:DoubleRCparameters}), and the comparison for $\Sigma_{RC}$ is listed for both the sample of 53 sightlines and that of all 87 sightlines investigated in this work. } 
\large
\centering 
\begin{tabular}{|l|cccc|} 
\hline\hline\hline 
  & $N_{\rm{Faint}}/$ & $\Sigma_{RC,53}$  & $\Sigma_{RC,87}$  & ${\Delta}I$ \\
  & $N_{\rm{Total}}$ & & & \\
 \hline\hline 
N12, $\alpha_{\rm{Bar}}=0^{\circ}$ & 0.11 & 0.12 & 0.12 & 0.08  \\
N12, $\alpha_{\rm{Bar}}=15^{\circ}$ & 0.07  & 0.12 & 0.12 & 0.07  \\
N12, $\alpha_{\rm{Bar}}=30^{\circ}$ & 0.05 & 0.11 & 0.13 & 0.04  \\
N12, $\alpha_{\rm{Bar}}=45^{\circ}$ & 0.09 & 0.11 & 0.17 & 0.06  \\
\hline
S10, $\alpha_{\rm{Bar}}=0^{\circ}$ & 0.09 & 0.12 & 0.13 & 0.10  \\
S10, $\alpha_{\rm{Bar}}=15^{\circ}$ & 0.06 & 0.15 & 0.15 & 0.09 \\
S10, $\alpha_{\rm{Bar}}=30^{\circ}$ & 0.05  & 0.11 & 0.15 & 0.05 \\
S10, $\alpha_{\rm{Bar}}=45^{\circ}$ & 0.05  & 0.12 & 0.18 & 0.05 \\
\hline\hline
\end{tabular}
\label{table:ModelDataComparison} 
\end{table}

\clearpage

\clearpage

\section{Summary}
\label{sec:Summary}

In this investigation, we have made use of reddening maps from \citet{2013ApJ...769...88N} to precisely characterise the luminosity function of red giant stars toward the high-latitude fields, with the resulting RGBB parameters $({\Delta}I_{RGBB}^{RC}=0.61,  f_{RGBB}^{RC}=0.16)$ being consistent  with those measured closer to the plane  $({\Delta}I_{RGBB}^{RC}=0.74,  f_{RGBB}^{RC}=0.20)$ if one assumes a metallicity difference of ${\Delta}$[M/H]$\approx0.22$ dex. The exponential slope of the RG luminosity function continuum can be fixed to $B=0.578$, at least for the directions probed in this work. The magnitude dispersion of the RC toward these sightlines (typically $\sim$0.22 mag for the brighter RC) is dominated by geometric dispersion, as evidenced by the fact that the ratio $\sigma_{RC,2} / \sigma_{RC,1} = 0.80$ can be explained purely by geometry. We have also found that for these sightlines, skew-Gaussians with $\alpha_{\rm{Skew}}=-1.50$ for the brighter RC and $\alpha_{\rm{Skew}}=+1.50$ for the fainter RC yield a superior fit to the luminosity function than standard Gaussians. 

The ratio of RC stars to RG stars is found to be much lower than predicted from stellar theory. We measure $EW_{RC} \approx 1.21 -0.164*(|b| -5)$. This can be reconciled with theory if 19\% and 27\% of stars near the brightness and colour of the RC are disk contaminants toward $|b|=5^{\circ}$ and $|b|=6.5^{\circ}$, respectively, similar to the estimate of $\sim 25\%$ derived by \citet{2013MNRAS.430..836N}. Alternatively, there may be a significant fraction of bulge stars in close binaries that lose extra mass as they ascend the RGB and thus never begin the core-helium-burning phase, as recently suggested by \citet{2014arXiv1406.6451C}.

The precise luminosity function and reddening maps allow us to map the directional-dependence of the brightness of the brighter RC, the brightness separation between the two RCs, the total surface density of RC stars on the sky,  and the fraction of stars in the brighter RC.  All of the observables show smooth behaviour toward $(-5.0^{\circ}  \lesssim l \lesssim 4.0^{\circ}  )$. Outside of that range, most observables yield erratic measurements. A notable exception being that of the total surface density of RC stars $\Sigma_{RC}$, an observable that is relatively impervious to systematic measurement errors. It falls off smoothly and rapidly outside that range.

In our comparison to N-body models, we have found that the case where the orientation angle $\alpha_{\rm{Bar}}=30^{\circ}$ is favoured by the model of   \citet{2012ApJ...756...22N} whereas $\alpha_{\rm{Bar}}=30,45^{\circ}$ are comparably favoured by the models of  \citet{2010ApJ...720L..72S}. As is expected, no perfect match is to be found. The biggest discrepancies between models and data are in the rate of decline of the combined surface density of red clump stars toward negative longitudes and of the brightness difference between the two RCs toward positive longitudes. 

The two models tested are, perhaps surprisingly, more distinct from the data as they are from one another in their predictions. Though some progress has been made in the literature in recent years in comparing observational data of the Galactic bulge to N-body models  \citep{2010ApJ...720L..72S,2011ApJ...734L..20M,2014ApJ...785L..17L,2012ApJ...757L...7L,2012ApJ...756...22N,2013A&A...555A..91V,2013MNRAS.432.2092N,2014MNRAS.438.3275G}, these discrepancies demonstrate that there remain significant issues to resolve, significant diversity in the allowed kinematic parameter space. We note that all models are internally-consistent dynamical pictures of what a galaxy could conceivably look like, but there are countless variables from the mass ratios of the different components to their initial angular momenta to the shape of the dark matter halo -- it would nearly be a miracle to achieve a perfect match to the Milky Way, which is one specific galaxy with its own distinct assembly history, in addition to possible undiagnosed systematics arising from the convolving of isochrones with N-body models to simulate how the data is observed. We have also not accounted for possible variations due to correlations between kinematics and chemistry, which are significant for the bulge stellar population  \citep{2008A&A...486..177Z,2010A&A...519A..77B,2012ApJ...756...22N,2013MNRAS.430..836N,2014A&A...563A..15B} and could distort the resulting luminosity function and interpretations thereof \citep{2014MNRAS.442.2075N}. Nevertheless, the Milky Way remains the only Galaxy that we can study in all six dimensions of kinematic phase space (plus chemistry) with great precisions, thus justifying attempts to delineate and decipher as many morphological components as accurately as possible, with corresponding explanation in  theoretically-understood dynamical phenomena. 

\section{Acknowledgments}
We thank Andrew Gould and Martin Asplund for helpful discussions. We thank the referee for a helpful report that improved the manuscript. 

DMN was primarily supported by the Australian Research Council grant FL110100012, and partially supported by the NSERC grant PGSD3-403304-2011 and the NSF grant AST-1103471. EA acknowledges financial support from the CNES (Centre National d'Etudes Spatiales - France) and from the People Programme  (Marie Curie Actions) of the European Union's Seventh Framework Programme FP7/2007-2013/ under REA grant agreement number PITN-GA-2011-289313 to the DAGAL network.  EA is thankful for HPC resources from GENCI- TGCC/CINES (Grants 2013 - x2013047098 and 2014 - x2014047098). 
JS acknowledges the support from the 973 Program of China under grants No. 2014CB845701, the National Natural Science Foundation of China under grants No. 11073037, 11333003, 11322326, and the Strategic Priority Research Program ``The Emergence of Cosmological Structures" (No. XDB09000000) of Chinese Academy of Sciences. ZYL is grateful for the support from Shanghai Yangfan Talent Youth Program (No. 14YF1407700).

The OGLE project has received funding from the European Research Council under the European Community's Seventh Framework Programme (FP7/2007-2013) / ERC grant agreement no. 246678 to AU. 

This work has made use of BaSTI web tools.

\newpage

\appendix


\begin{thebibliography}{}
\bibitem[Athanassoula(2003)]{2003MNRAS.341.1179A} Athanassoula, E.\ 2003, 
\mnras, 341, 1179 
\bibitem[Athanassoula(2005)]{2005MNRAS.358.1477A} Athanassoula, E.\ 2005, 
\mnras, 358, 1477 
\bibitem[Babusiaux 
\& Gilmore(2005)]{2005MNRAS.358.1309B} Babusiaux, C., \& Gilmore, G.\ 2005, \mnras, 358, 1309
\bibitem[Babusiaux et 
al.(2010)]{2010A&A...519A..77B} Babusiaux, C., G{\'o}mez, A., Hill, V., et al.\ 2010, \aap, 519, A77 
\bibitem[Babusiaux et 
al.(2014)]{2014A&A...563A..15B} Babusiaux, C., Katz, D., Hill, V., et al.\ 2014, \aap, 563, A15 
\bibitem[Bjork 
\& Chaboyer(2006)]{2006ApJ...641.1102B} Bjork, S.~R., \& Chaboyer, B.\ 2006, \apj, 641, 1102 
\bibitem[Bureau et al.(2006)]{2006MNRAS.370..753B} Bureau, M., Aronica, G., 
Athanassoula, E., et al.\ 2006, \mnras, 370, 753 
\bibitem[Cassisi 
\& Salaris(1997)]{1997MNRAS.285..593C} Cassisi, S., \& Salaris, M.\ 1997, \mnras, 285, 593 
\bibitem[Calamida et al.(2014)]{2014arXiv1406.6451C} Calamida, A., Sahu, 
K.~C., Anderson, J., et al.\ 2014, arXiv:1406.6451 
\bibitem[Cao et al.(2013)]{2013MNRAS.434..595C} Cao, L., Mao, S., Nataf, 
D., Rattenbury, N.~J., \& Gould, A.\ 2013, \mnras, 434, 595 
\bibitem[Chatzopoulos et al.(2014)]{2014arXiv1403.5266C} Chatzopoulos, S., 
Fritz, T., Gerhard, O., et al.\ 2014, arXiv:1403.5266 
\bibitem[{{Cordier} et~al.(2007){Cordier}, {Pietrinferni}, {Cassisi} \&
  {Salaris}}]{2007AJ....133..468C}
{Cordier} D., {Pietrinferni} A., {Cassisi} S., {Salaris} M., 2007, \aj, 133,
  468
\bibitem[Dwek et al.(1995)]{1995ApJ...445..716D} Dwek, E., Arendt, R.~G., 
Hauser, M.~G., et al.\ 1995, \apj, 445, 716 
\bibitem[Freeman et al.(2013)]{2013MNRAS.428.3660F} Freeman, K., Ness, M., 
Wylie-de-Boer, E., et al.\ 2013, \mnras, 428, 3660 
\bibitem[Gardner et al.(2014)]{2014MNRAS.438.3275G} Gardner, E., 
Debattista, V.~P., Robin, A.~C., V{\'a}squez, S., 
\& Zoccali, M.\ 2014, \mnras, 438, 3275 
\bibitem[Gillessen et al.(2009)]{2009ApJ...692.1075G} Gillessen, S., 
Eisenhauer, F., Trippe, S., et al.\ 2009, \apj, 692, 1075 
\bibitem[Gonzalez et 
al.(2012)]{2012A&A...543A..13G} Gonzalez, O.~A., Rejkuba, M., Zoccali, M., et al.\ 2012, \aap, 543, A13 
\bibitem[Howard et al.(2008)]{2008ApJ...688.1060H} Howard, C.~D., Rich, 
R.~M., Reitzel, D.~B., et al.\ 2008, \apj, 688, 1060 
\bibitem[Kunder et al.(2012)]{2012AJ....143...57K} Kunder, A., Koch, A., 
Rich, R.~M., et al.\ 2012, \aj, 143, 57 
\bibitem[Laurikainen et al.(2013)]{2013MNRAS.430.3489L} Laurikainen, E., 
Salo, H., Athanassoula, E., et al.\ 2013, \mnras, 430, 3489 
\bibitem[Laurikainen et al.(2014)]{2014arXiv1406.1418L} Laurikainen, E., 
Salo, H., Athanassoula, E., Bosma, A., 
\& Herrera-Endoqui, M.\ 2014, arXiv:1406.1418 
\bibitem[Li 
\& Shen(2012)]{2012ApJ...757L...7L} Li, Z.-Y., \& Shen, J.\ 2012, \apjl, 757, L7 
\bibitem[Li et al.(2014)]{2014ApJ...785L..17L} Li, Z.-Y., Shen, J., Rich, 
R.~M., Kunder, A., \& Mao, S.\ 2014, \apjl, 785, L17 
\bibitem[Martinez-Valpuesta 
\& Gerhard(2011)]{2011ApJ...734L..20M} Martinez-Valpuesta, I., \& Gerhard, O.\ 2011, \apjl, 734, L20 
\bibitem[McWilliam 
\& Zoccali(2010)]{2010ApJ...724.1491M} McWilliam, A., \& Zoccali, M.\ 2010, \apj, 724, 1491 
\bibitem[Nataf et al.(2010)]{2010ApJ...721L..28N} Nataf, D.~M., Udalski, 
A., Gould, A., Fouqu{\'e}, P., \& Stanek, K.~Z.\ 2010, \apjl, 721, L28 
\bibitem[Nataf et al.(2011)]{2011ApJ...730..118N} Nataf, D.~M., Udalski, 
A., Gould, A., \& Pinsonneault, M.~H.\ 2011, \apj, 730, 118 
\bibitem[Nataf et al.(2013a)]{2013ApJ...766...77N} Nataf, D.~M., Gould, 
A.~P., Pinsonneault, M.~H., \& Udalski, A.\ 2013a, \apj, 766, 77 
\bibitem[Nataf et al.(2013b)]{2013ApJ...769...88N} Nataf, D.~M., Gould, A., 
Fouqu{\'e}, P., et al.\ 2013b, \apj, 769, 88 
\bibitem[Nataf et al.(2014)]{2014MNRAS.442.2075N} Nataf, D.~M., Cassisi, 
S., \& Athanassoula, E.\ 2014, \mnras, 442, 2075 
\bibitem[Ness et al.(2012)]{2012ApJ...756...22N} Ness, M., Freeman, K., 
Athanassoula, E., et al.\ 2012, \apj, 756, 22 
\bibitem[Ness et al.(2013)]{2013MNRAS.430..836N} Ness, M., Freeman, K., 
Athanassoula, E., et al.\ 2013, \mnras, 430, 836 
\bibitem[Ness et al.(2013)]{2013MNRAS.432.2092N} Ness, M., Freeman, K., 
Athanassoula, E., et al.\ 2013, \mnras, 432, 2092 
\bibitem[Patsis et al.(2002)]{2002MNRAS.337..578P} Patsis, P.~A., Skokos, 
C., \& Athanassoula, E.\ 2002, \mnras, 337, 578 
\bibitem[Pietrinferni et al.(2004)]{2004ApJ...612..168P} Pietrinferni, A., 
Cassisi, S., Salaris, M., \& Castelli, F.\ 2004, \apj, 612, 168 
\bibitem[Poleski et al.(2013)]{2013ApJ...776...76P} Poleski, R., Udalski, 
A., Gould, A., et al.\ 2013, \apj, 776, 76 
\bibitem[Romero-G{\'o}mez et al.(2011)]{2011MNRAS.418.1176R} 
Romero-G{\'o}mez, M., Athanassoula, E., Antoja, T., 
\& Figueras, F.\ 2011, \mnras, 418, 1176 
\bibitem[Salaris 
\& Girardi(2002)]{2002MNRAS.337..332S} Salaris, M., \& Girardi, L.\ 2002, \mnras, 337, 332 
\bibitem[Saito et 
al.(2012)]{2012A&A...537A.107S} Saito, R.~K., Hempel, M., Minniti, D., et al.\ 2012, \aap, 537, A107 
\bibitem[Salpeter(1955)]{1955ApJ...121..161S} Salpeter, E.~E.\ 1955, \apj, 
121, 161 
\bibitem[Sch{\"o}nrich(2012)]{2012MNRAS.427..274S} Sch{\"o}nrich, R.\ 2012, 
\mnras, 427, 274 
\bibitem[Shen et al.(2010)]{2010ApJ...720L..72S} Shen, J., Rich, R.~M., 
Kormendy, J., et al.\ 2010, \apjl, 720, L72 
\bibitem[Shen(2014)]{2014IAUS..298..201S} Shen, J.\ 2014, IAU Symposium, 
298, 201 
\bibitem[Skokos et al.(2002)]{2002MNRAS.333..861S} Skokos, C., Patsis, 
P.~A., \& Athanassoula, E.\ 2002, \mnras, 333, 861 
\bibitem[Skrutskie et al.(2006)]{2006AJ....131.1163S} Skrutskie, M.~F., 
Cutri, R.~M., Stiening, R., et al.\ 2006, \aj, 131, 1163 
\bibitem[Stanek et al.(1994)]{1994ApJ...429L..73S} Stanek, K.~Z., Mateo, 
M., Udalski, A., et al.\ 1994, \apjl, 429, L73 
\bibitem[Stanek et al.(1997)]{1997ApJ...477..163S} Stanek, K.~Z., Udalski, 
A., Szyma{\'n}ski, M., et al.\ 1997, \apj, 477, 163 
\bibitem[Szyma{\'n}ski et al.(2011)]{2011AcA....61...83S} Szyma{\'n}ski, 
M.~K., Udalski, A., Soszy{\'n}ski, I., et al.\ 2011, \actaa, 61, 83 
\bibitem[Udalski et al.(2002)]{2002AcA....52..217U} Udalski, A., Szymanski, 
M., Kubiak, M., et al.\ 2002, \actaa, 52, 217 
\bibitem[Udalski(2003)]{2003AcA....53..291U} Udalski, A.\ 2003, Acta 
Astronomica, 53, 291 
\bibitem[Udalski et al.(2008)]{2008AcA....58...69U} Udalski, A., Szyma{\'n}ski, 
M.~K., Soszy{\'n}ski, I., \& Poleski, R.\ 2008, Acta Astronomica, 58, 69 
\bibitem[Uttenthaler et 
al.(2012)]{2012A&A...546A..57U} Uttenthaler, S., Schultheis, M., Nataf, D.~M., et al.\ 2012, \aap, 546, A57 
\bibitem[V{\'a}squez et 
al.(2013)]{2013A&A...555A..91V} V{\'a}squez, S., Zoccali, M., Hill, V., et al.\ 2013, \aap, 555, A91 
\bibitem[Wegg 
\& Gerhard(2013)]{2013MNRAS.435.1874W} Wegg, C., \& Gerhard, O.\ 2013, \mnras, 435, 1874  
\bibitem[Zoccali et 
al.(2008)]{2008A&A...486..177Z} Zoccali, M., Hill, V., Lecureur, A., et al.\ 2008, \aap, 486, 177 
\end{thebibliography}
\end{document}